\documentclass[acmsmall,screen,natbib,nonacm]{acmart}
\usepackage[utf8]{inputenc}
\usepackage[frozencache]{minted}

\title{Transform Dialect Tutorial}
\author{Oleksandr Zinenko}
\orcid{0000-0003-1978-0222}
\affiliation{
  \institution{Google DeepMind}
  \city{Paris}
  \country{France}
}
\email{zinenko@google.com}
\date{April 2024}

\newmintinline[mlircode]{c}{fontsize=\footnotesize}
\newmintinline[cppcode]{cpp}{fontsize=\footnotesize}

\makeatletter
\AtBeginEnvironment{minted}{\dontdofcolorbox}
\def\dontdofcolorbox{\renewcommand\fcolorbox[4][]{##4}}
\makeatother

\newenvironment{cmakeblock}{
\VerbatimEnvironment
\begin{minted}[
frame=lines,
fontsize=\footnotesize
]{cmake}}
{
\end{minted}
}

\newenvironment{cppblock}{
\VerbatimEnvironment
\begin{minted}[
frame=lines,
fontsize=\footnotesize
]{cpp}}
{
\end{minted}
}

\newenvironment{mlirblock}{
\VerbatimEnvironment
\begin{minted}[
frame=lines,
fontsize=\footnotesize
]{'mlirlexer.py:MlirLexer -x'}}
{
\end{minted}
}

\newenvironment{shellblock}{
\VerbatimEnvironment
\begin{minted}[
breaklines,
frame=lines,
fontsize=\footnotesize
]{shell}}
{
\end{minted}
}

\newenvironment{tdblock}{
\VerbatimEnvironment
\begin{minted}[
breaklines,
frame=lines,
fontsize=\footnotesize
]{'tdlexer.py:TdLexer -x'}}
{
\end{minted}
}

\newenvironment{pyblock}{
\VerbatimEnvironment
\begin{minted}[
breaklines,
frame=lines,
fontsize=\footnotesize
]{python}}
{
\end{minted}
}

\newenvironment{asmblock}{
\VerbatimEnvironment
\begin{minted}[
breaklines,
frame=lines,
fontsize=\footnotesize
]{asm}}
{
\end{minted}
}

\begin{abstract}
Transform Dialect in MLIR~\cite{mlir} provides operations that can be used to control
transformation of the Intermediate Representation (IR) using a different
portion of the IR. It refers to the IR being transformed as payload IR, and to
the IR guiding the transformation as transform IR.

The main use case for this dialect is orchestrating fine-grain transformations
on individual IR objects (operations or values) or sets thereof. For example,
it may involve finding loop-like operations with specific properties (e.g.,
large size) in the payload IR, applying loop tiling to those and only those
operations, and then applying loop unrolling to the inner loops produced by the
previous transformations. As such, it is not intended as a replacement for the
pass infrastructure, nor for the pattern rewriting infrastructure. In the most
common case, the transform IR will be processed and applied to the payload IR
by a pass. Transformations expressed by the Transform dialect may be
implemented using the pattern infrastructure or any other relevant MLIR
component.

The rest of this document explains the main concepts and usage scenario of the
MLIR Transform Dialect combined with structured~\cite{structured} operations.
\end{abstract}

\begin{document}

\maketitle

\tableofcontents

\section{A Primer on ``Structured'' Linalg Operations}

Before starting the tutorial on the Transform dialect, let us take a brief look
at the concept of Structured operations and its implementation in the Linalg
dialect. Note that the Transform dialect does not require Structured operations
and vice versa. The two co-evolved at the beginning of the Transform dialect,
which makes the subset of transformations for Structured operations the most
mature and most suitable for the tutorial. If you are already familiar with
this concept, skip to the next section.

Structured code generation intends to preserve the structure of the computation
for as long as necessary to enable transformations, up to and including the
design of IR abstractions that support specific transformations.

\subsection{Uniform Elementwise Extension}

Consider a simple scalar arithmetic addition operation in MLIR, which maps
directly to a machine instruction on most architectures that support floating
point operations:

\begin{mlirblock}
%2 = arith.addf %0, %1 : f32
\end{mlirblock}

This operation can be easily extended to uniformly apply to elements of a 1D
vector, which is also often available as an instruction of vector machines:

\begin{mlirblock}
%2 = arith.addf %0, %1 : vector<8xf32>
\end{mlirblock}

Only a few modern instruction sets offer instructions for two- or
more-dimensional vectors. In MLIR, however, it is possible to transparently
extend the uniform elementwise application to vectors of arbitrary rank.

\begin{mlirblock}
%2 = arith.addf %0, %1 : vector<8x4xf32>
%5 = arith.addf %3, %4 : vector<2x2x2x2x2x2x2xf32>
\end{mlirblock}

As you can notice, MLIR’s arithmetic operations on vectors preserve the
structure of uniform elementwise application. This structure can be leveraged
by the compiler, for example, to produce smaller-rank operations available on
the target or to fuse multiplication and addition when such a fused instruction
is available (which becomes complicated when there are a hundred of
multiplications followed by a hundred of additions).

\subsection{Reduction}

Sometimes it is necessary to add elements of a vector to obtain a scalar. Some
platforms provide specific instructions for this operation, some others provide
ones that can be combined to achieve the desired effect, such as addition of
adjacent elements and element shuffle.

The Vector dialect in MLIR defines an operation to explicitly denote a
within-vector reduction:

\begin{mlirblock}
%0 = vector.reduction <add>, %0 : vector<8xf32> into f32
\end{mlirblock}

When no support is available, such an operation can be transformed into a loop:

\begin{mlirblock}
%c0 = arith.constant 0 : index
%c1 = arith.constant 1 : index
%c8 = arith.constant 8 : index
%init = arith.constant 0.0 : f32
%result = scf.for %i = %c0 to %c8 step %c1 iter_args(%partial = %init) -> (f32) {
  %element = vector.extractelement %0[%i : index] : vector<8xf32>
  %updated = arith.addf %partial, %element : f32
  scf.yield %updated : f32
}
\end{mlirblock}

Even when special instructions are available, it may still be desirable to use
the loop form (with unrolling), depending on instruction latency and register
pressure. Preserving the structure of the operation as a single reduction gives
the compiler an understanding that a within-vector reduction is performed and,
therefore, a choice in implementation.

\subsection{Contraction}

Contraction is a generalization of reduction that multiplies elements from two
vectors before adding them up. A simple ``add'' reduction can be thought of as a
contraction where one of the vectors contains \mintinline{text}{1.0}, the
neutral element of multiplication. Contractions offer even more flexibility to
the compiler, and are represented by a dedicated operation in MLIR:

\begin{mlirblock}
// Neutral initializer for the addition.
%init  = arith.constant 0.0 : f32
// Neutral element of multiplication.
%ones = arith.constant dense<1.0> : vector<8xf32>
// Actual contraction.
%result = vector.contract {
  indexing_maps = [affine_map<(i) -> (i)>,
                   affine_map<(i) -> (i)>,
                   affine_map<(i) -> ()>],
  iterator_types = ["reduction"]
} %0, %ones, %init : vector<8xf32>, vector<8xf32> into f32
\end{mlirblock}

Note the \mlircode{affine_map} expressions indicating how vector elements are
indexed. Their meaning is perhaps most evident when writing the loop form in
pseudo-code equivalent to this contraction:

\begin{pyblock}
for i in 0 to 8:
  init += p0[i] * ones[i]
\end{pyblock}

where both \mlircode{
\mlircode{i}, as noted on the right-hand side of the corresponding affine map,
\mlircode{(i) -> (i)}, and \mlircode{
right-hand side of its affine map, \mlircode{(i) -> ()}.

Similarly to uniform elementwise extension, MLIR vector contractions are not
limited to 1D cases. In the 2D+ case, one can additionally specify which of the
vector dimensions are being reduced and which ones are being preserved. This
can be achieved by using the \mlircode{iterator_types} attribute that specifies, for
each dimension, whether it is being reduced (\mlircode{"reduction"}) or preserved
(\mlircode{"parallel"}). Consider the following 3D contraction that encodes a
matrix-matrix multiplication:

\begin{mlirblock}
%result = vector.contract {
  indexing_maps = [affine_map<(i, j, k) -> (i, k)>,
                   affine_map<(i, j, k) -> (k, j)>,
                   affine_map<(i, j, k) -> (i, j)>],
  iterator_types = ["parallel", "parallel", "reduction"]
} %lhs, %rhs, %init: vector<8x10xf32>, vector<10x16xf32> into vector<8x16xf32>
\end{mlirblock}

Looking at the indexing maps, it is easy to recognize the loop form:

\begin{pyblock}
for i in 0 to 8:
  for j in 0 to 16:
    for k in 0 to 10:
      init[i, j] += lhs[i, k] * rhs[k, j]
\end{pyblock}

Preserving this higher-level structure of a contraction makes it significantly
easier for the compiler to recognize operations such as matrix multiplications
and dot products and gives it freedom to produce lower-level operations that
leverage most advanced instructions or even pre-generated microkernels.

\subsection{Generic Operation on Memory}

Until now, we have been considering operations on vectors stored in virtual
registers. A similar contraction abstraction can be defined in memory:

\begin{mlirblock}
linalg.generic {
  indexing_maps = [affine_map<(i, j, k) -> (i, k)>,
                   affine_map<(i, j, k) -> (k, j)>,
                   affine_map<(i, j, k) -> (i, j)>],
  iterator_types = ["parallel", "parallel", "reduction"]
} ins(%lhs, %rhs : memref<8x10xf32>, memref<10x16xf32>)
  outs(%init : memref<8x16xf32>) {
^bb0(%lhs_one: f32, %rhs_one: f32, %init_one: f32):
  %0 = arith.mulf %lhs_one, %rhs_one : f32
  %1 = arith.addf %init_one, %0 : f32
  linalg.yield %1 : f32
}
\end{mlirblock}

This looks more complicated, so let us unpack. The \mlircode{indexing_maps} and \mlircode{iterator_types} are \emph{exactly} the same as we have seen above for vector contractions. The operands are now split into two lists:

\begin{itemize}
\item \mlircode{in} operands containing the buffers that are being only read by the operation;
\item \mlircode{out} operands that are being read and updated by the operation.
\end{itemize}

This separation wasn’t necessary on vectors because, in MLIR, vectors are
read-only (SSA or functional form) and operations mutating a vector are in fact
producing a new one instead.

Furthermore, the operation now contains a region that explicitly specifies the
multiplication and the addition operations that were implicit in the
contraction. Block arguments in the region correspond to individual elements
read from the buffer: the first two correspond to the \mlircode{in} operands and the
last one corresponds to the \mlircode{out} operand. The value yielded from the region is
“written” to the \mlircode{out} operand and is available as the last block argument for
future executions of the region. Note that the order in which the region is
executed for various tuples of elements read from the buffers is not specified,
and the write to the \mlircode{out} buffer is written as a whole at the end of the
operation.

\subsection{``Loop'' Fusion}

Since the region of the \mlircode{linalg.generic} operation can contain
arbitrarily many operations, we can use it to express “fusion” of the implicit
loops by simply having more operations chained in the region. For example, the
common machine learning rectified linear unit layer (ReLU), which can be
defined as \mlircode{relu(x) = max(0, x)}, can be defined be expressed using
the “compare-and-select” idiom in one \mlircode{linalg.generic} operation,
without the temporary buffer for the comparison result and without repeating
the outer operation:

\begin{mlirblock}
linalg.generic {
  indexing_maps [affine_map<(i) -> (i)>, affine_map<(i) -> (i)>],
  iterator_types = ["parallel"]
} ins(%in : memref<?xf32>) outs(%out : memref<?xf32>) {
^bb0(%in_one : f32, %out_one : f32):
  %c0 = arith.constant 0.0 : f32
  %0 = arith.cmpf ogt %in_one, %c0 : f32
  %1 = arith.select %0, %in_one, %c0 : f32
  linalg.yield %1 : f32
}
\end{mlirblock}

Such operations can be converted to loops or lowered into vector forms after
splitting into multiple operations, each of which maps to a Vector dialect
primitive. This modeling, again, gives the compiler more choice in selecting
the code generation strategy.

\subsection{Generic Operation on Tensors}

Let us take one last step up on the abstraction ladder. MLIR provides a tensor
abstraction that makes it easy for the compiler to reason about
multidimensional yet regular data without having to solve complex problems such
as alias analysis and dependency satisfaction, which would be necessary on
multidimensional buffers. The tensor abstraction is very similar to the vector
abstraction (major differences include the availability of unranked tensors,
tensor layouts, and vectors being usable as elemental types of tensors but not
of other vectors). Tensors are read-only, and operations updating a tensor
produce a new tensor.

The \mlircode{linalg.generic} operation from above can lifted to operate on tensors instead of buffers:

\begin{mlirblock}
%result = linalg.generic {
  indexing_maps = [affine_map<(i, j, k) -> (i, k)>,
                   affine_map<(i, j, k) -> (k, j)>,
                   affine_map<(i, j, k) -> (i, j)>],
  iterator_types = ["parallel", "parallel", "reduction"]
} ins(%lhs, %rhs : tensor<8x10xf32>,tensor<10x16xf32>)
  outs(%init :tensor<8x16xf32>) {
^bb0(%lhs_one: f32, %rhs_one: f32, %init_one: f32):
  %0 = arith.mulf %lhs_one, %rhs_one : f32
  %1 = arith.addf %init_one, %0 : f32
  linalg.yield %1 : f32
} -> tensor<8x16xf32>
\end{mlirblock}

As you can notice, most components of this operation remain identical to its
buffer version. It has been specifically designed this way. The main
difference, beside the operand types, is that the operation now produces a new
result instead of updating the \mlircode{out} buffer. The \mlircode{out}
operand is used only as the initialization value.

If the \mlircode{linalg.generic} operation had existed on vectors, it would
have had the exact same structure.

\subsection{Tiling and Loop Materialization}

At this level of abstraction, it becomes easy for the compiler to perform more
advanced transformations usually required for high-performance code generation,
such as \href{https://en.wikipedia.org/wiki/Loop_nest_optimization}{tiling}.
Tiling, in general, can be seen as partitioning the iteration space into
smaller parts, or tiles, so that the data required by each part fits into a
level of cache for example. The order in which tiles are executed must preserve
the original data dependencies.

In the case of \mlircode{linalg.generic} operations, the iteration space is
implicit and is defined by the shape of the operands. Therefore, a tile can be
expressed by performing the \emph{same} operation on a subset (slice) of the
original data. Since the order in which the body of \mlircode{linalg.generic}
is applied to different tuples of the input elements is unspecified, tiles can
be executed in any order, without the need for dependence analysis. In order to
control the execution of different tiles, the implementation of tiling produces
loops. Thus tiling \mlircode{linalg.generic} operations can also be seen as
materializing the loops that have been implicit until now.

For example, tiling the matrix multiplication presented above with tile sizes
\mlircode{(2, 8)}, we obtain a loop nest around a \mlircode{linalg.generic}
expressing the same operation on a \mlircode{2x8} tensor.

\begin{mlirblock}
// A special "multi-for" loop that supports tensor-insertion semantics
// as opposed to implicit updates. The resulting 8x16 tensor will be produced
// by this loop.
// The trip count of iterators is computed dividing the original tensor size,
// 8x16, by the tile size, 2x8, to obtain 4x2.
// When tensor sizes are dynamic, the trip count computation is emitted as IR
// and is being computed at runtime.
%0 = scf.forall (%i, %j) in (4, 2)
     shared_outs(%shared = %init) -> (tensor<8x16xf32>) {

  // Scale the loop induction variables by the tile sizes.
  %3 = affine.apply affine_map<(d0) -> (d0 * 2)>(%i)
  %4 = affine.apply affine_map<(d0) -> (d0 * 8)>(%j)

  // Take slices of inputs and outputs. Only the "i" and "j" dimensions are sliced.
  %lhs_slice = tensor.extract_slice %lhs[%3, 0] [2, 10] [1, 1]
             : tensor<8x10xf32> to tensor<2x10xf32>
  %rhs_slice = tensor.extract_slice %rhs[0, %4] [10, 8] [1, 1]
             : tensor<10x16xf32> to tensor<10x8xf32>
  %result_slice = tensor.extract_slice %shared[%3, %4] [2, 8] [1, 1]
                : tensor<8x16xf32> to tensor<2x8xf32>

  // This is exactly the same operation as before, but now operating on smaller
  // slices of data.
  %partial =  linalg.generic {
  indexing_maps = [affine_map<(i, j, k) -> (i, k)>,
                   affine_map<(i, j, k) -> (k, j)>,
                   affine_map<(i, j, k) -> (i, j)>],
  iterator_types = ["parallel", "parallel", "reduction"]
  } ins(%lhs_slice, %rhs_slice : tensor<2x10xf32>, tensor<10x8xf32>)
    outs(%result_slice : tensor<2x8xf32>) -> tensor<2x8xf32> {
  ^bb0(%lhs_one: f32, %rhs_one: f32, %init_one: f32):
    %0 = arith.mulf %lhs_one, %rhs_one : f32
    %1 = arith.addf %init_one, %0 : f32
    linalg.yield %1 : f32
  } : tensor<2x8xf32>

  // Terminator for the loop with tensor-insertion semantics. Inserts a slice
  // into a larger tensor, potentially in parallel.
  scf.forall.in_parallel {
    tensor.parallel_insert_slice %partial into %shared[%3, %4] [2, 8] [1, 1]
        : tensor<2x8xf32> into tensor<8x16xf32>
  }
}
\end{mlirblock}

\subsection{Producer/Consumer Fusion and Rematerialization}

After materializing loops with tiling, another key code generation
transformation becomes simple – fusion. Unlike loop fusion, the Structured
operations approach allows for producer/consumer fusion even when the
(implicit) iteration spaces of the operations do not match. Given an high-level
structured operation on tensors, such as \mlircode{linalg.generic}, one can follow
use-def chains to identify:

\begin{enumerate}
  \item the subset (slice) of the operand that is used by the tile, and
  \item the tensor-level structured operation producing the whole tensor that
	is being sliced.
\end{enumerate}

By inverting the \mlircode{indexing_map} and applying it to the set of elements accessed through the slice, we can compute the part of the iteration space of the operation defining the full tensor necessary to compute the tile. Thus fusion boils down to replacing the \mlircode{tensor.extract_slice} operation with the tile of the \mlircode{linalg.generic} producing the original operand.

Let us assume that the matrix multiplication operation is followed by another operation that multiplies each element of the resulting matrix with itself. This trailing elementwise operation has a 2D iteration space, unlike the 3D one in matrix multiplication. Nevertheless, it is possible to tile the trailing operation and then fuse the producer of its operand, the matmul, into the loop generated by tiling. The untiled dimension will be used in its entirety.

\begin{mlirblock}
// Same loop as before.
%0 = scf.forall (%i, %j) in (4, 2)
     shared_outs(%shared = %init)
     -> (tensor<8x16xf32>, tensor<8x16xf32>) {
  // Scale the loop induction variables by the tile sizes.
  %1 = affine.apply affine_map<(d0) -> (d0 * 2)>(%i)
  %2 = affine.apply affine_map<(d0) -> (d0 * 8)>(%j)

  // Take slices of inputs and outputs. Only the "i" and "j" dimensions are sliced.
  %lhs_slice = tensor.extract_slice %lhs[%1, 0] [2, 10] [1, 1]
             : tensor<8x10xf32> to tensor<2x10xf32>
  %rhs_slice = tensor.extract_slice %rhs[0, %2] [10, 8] [1, 1]
             : tensor<10x16xf32> to tensor<10x8xf32>
  %result_slice = tensor.extract_slice %result[%1, %2] [2, 8] [1, 1]
                : tensor<8x16xf32> to tensor<2x8xf32>

  // This is exactly the same matmul slice as before. It replaces the slice
  // extraction for the generic operation below.
  %partial = linalg.generic {
    indexing_maps = [affine_map<(i, j, k) -> (i, k)>,
                     affine_map<(i, j, k) -> (k, j)>,
                     affine_map<(i, j, k) -> (i, j)>],
    iterator_types = ["parallel", "parallel", "reduction"]
  } ins(%lhs_slice, %rhs_slice : tensor<2x10xf32>, tensor<10x8xf32>)
   outs(%result_slice : tensor<2x8xf32>) {
  ^bb0(%lhs_one: f32, %rhs_one: f32, %init_one: f32):
    %5 = arith.mulf %lhs_one, %rhs_one : f32
    %6 = arith.addf %init_one, %5 : f32
    linalg.yield %6 : f32
  } -> tensor<2x8xf32>

  // Take the slice of the final result. Note that we don't need to take
  // the slice of the operand because the matmul operation above computes
  // it in-place.
  %shared_slice = tensor.extract_slice %shared[%1, %2] [2, 8] [1, 1]
                : tensor<8x16xf32> to tensor<2x8xf32>

  // The elementwise operation that we tiled.
  %elemwise = linalg.generic {
    indexing_maps = [affine_map<(i, j) -> (i, j)>,
                     affine_map<(i, j) -> (i, j)>],
    iterator_types = ["parallel", "parallel"]
  } ins(%partial : tensor<2x8xf32>)
   outs(%shared_slice : tensor<2x8xf32>) {
  ^bb0(%in: f32, %out: f32):
    %5 = arith.mulf %in, %in : f32
    linalg.yield %5 : f32
  } -> tensor<2x8xf32>

  // Terminator for the loop with tensor-insertion semantics. Inserts a slice
  // into a larger tensor, potentially in parallel.
  scf.forall.in_parallel {
    tensor.parallel_insert_slice %elemwise into %shared[%1, %2] [2, 8] [1, 1]
        : tensor<2x8xf32> into tensor<8x16xf32>
  }
}
\end{mlirblock}

This process may result in some elements in the operand tensors being
(re)computed on every iteration of the loop. This is also known as
\emph{rematerialization} and expresses the tradeoff between performing
redundant computations or storing their result in (slow) memory.

\subsection{Shorthand ``Named'' Forms of Linalg Ops}

Linalg provides a set of predefined operations for common cases such as matrix
multiplication, dot product, convolution, etc. These operations are equivalent
to the \mlircode{generic} ones but spare the need to spell out the access patterns and
the bodies. For example, matrix multiplication is simply:

\begin{mlirblock}
%matmul = linalg.matmul ins(%lhs, %rhs: tensor<8x10xf32>, tensor<10x16xf32>)
                        outs(%init: tensor<8x10xf32xf32>) -> tensor<8x16xf32>
\end{mlirblock}

\section{Combining Existing Transformations}
\subsection{Introduction}
The Transform dialect allows one to precisely target transformations at specific operations in the IR and to chain them, that is to apply a transformation to operations produced by the previous transformation. To achieve this, transformations are expressed as other operations in the IR. We call these the IR containing these operations transform IR. And we call the IR that is being transformed payload IR.

Transform IR operations operate on values that may be associated with payload IR operations, values or attributes. We call the first two kinds of values operation and value handles, respectively. We call the last kind of values parameters.

The application of transform IR always starts from one top-level operation. In the C++ API, this operation is passed to the \cppcode{applyTransforms} function. This top-level operation specifies if other transformations should be performed and how. The most common top-level operation, \mlircode{transform.named_sequence} merely applies other transform operations listed in its body one after the other, similarly to a function or a macro.

Let us illustrate this with a simple sequence of transformations on the common ``fully connected + bias + ReLU'' ML layer, which boils down to performing a matrix multiplication, followed by an (elementwise) matrix addition and taking an elementwise maximum with 0. This can be expressed using the following IR:

\begin{mlirblock}
func.func @fc_relu(%lhs: tensor<512x512xf32>, %rhs: tensor<512x512xf32>,
                   %bias: tensor<512x512xf32>, %output: tensor<512x512xf32>)
                   -> tensor<512x512xf32> {
  // Matrix-matrix multiplication.
  %matmul = linalg.matmul ins(%lhs, %rhs: tensor<512x512xf32>, tensor<512x512xf32>)
                          outs(%output: tensor<512x512xf32>) -> tensor<512x512xf32>

  // Elementwise addition.
  %biased = linalg.elemwise_binary { fun = #linalg.binary_fn<add> }
    ins(%matmul, %bias : tensor<512x512xf32>, tensor<512x512xf32>)
    outs(%output : tensor<512x512xf32>) -> tensor<512x512xf32>

  // Elementwise max with 0 (ReLU).
  %c0f = arith.constant 0.0 : f32
  %relued = linalg.elemwise_binary { fun = #linalg.binary_fn<max_signed> }
    ins(%biased, %c0f : tensor<512x512xf32>, f32)
    outs(%output : tensor<512x512xf32>) -> tensor<512x512xf32>
  func.return %relued : tensor<512x512xf32>
}
\end{mlirblock}

\subsection{Top-Level Sequence Operation}
For performance reasons, we would like to tile and fuse these operations to exploit cache locality. This is a sequence of transformations that need to be performed one after another, so we naturally start with the corresponding top-level transform operation.

\begin{mlirblock}
module attributes {transform.with_named_sequence} {
  transform.named_sequence @__transform_main(
      %arg0: !transform.any_op,
      %arg1: !transform.op<"linalg.matmul">,
      %arg2: !transform.op<"linalg.elemwise_binary">):
    transform.yield
  }
}
\end{mlirblock}
There are several aspects worth noticing in this operation.

Its special name, \mlircode{@__transform_main} and the first argument are mandated by the interpreter pass, similarly to how the entry point of C programs needs to be called main and may have the \cppcode{int (int argc, char** argv)} signature. This argument will be associated with the top-level payload operation, most often the operation that the pass is applied to. Note that none of this is required when applying the transformation programmatically via \cppcode{applyTransforms} or \cppcode{applyNamedSequence}.

The remaining entry block arguments are optional and can be associated with payload attributes, operations or values that are useful in the sequence. These are also specified when calling \cppcode{applyTransforms}. In our case, we are interested in the matrix multiplication and elementwise operations that we are going to tile and fuse.

All value handles have Transform dialect types. These types specify certain properties of the payload IR entities associated with them. In this example, \mlircode{transform.any_op} indicates that the handle is associated with arbitrary payload operations. On the contrary, \mlircode{transform.op<"X">} indicates that the handle is associated only with payload operations of kind X. These constraints are verified when the handle/payload association is created. For entry block arguments of top-level transform operations, this happens early in the \cppcode{applyTransforms} function. If the constraints are not satisfied, the transform application fails and produces diagnostics for the user.

Finally, the operation is wrapped in a module with the \mlircode{transform.with_named_sequence} attribute that triggers all necessary verifications if multiple named sequences exist.

\subsection{Failure Propagation}
The Transform dialect infrastructure has a particular mechanism for handling diagnostics that supports recoverable errors. It is best understood by considering the (unnamed) sequence operation that has a mandatory attribute specifying the failure propagation mode. There are two options:

\mlircode{"propagate"} makes the sequence transformation fail if any of the nested transformation fails;
\mlircode{"suppress"} makes the sequence succeed even if one of the nested transformations fails, but without attempting to perform the transformations following the failed one in the sequence.
This latter allows the transformation script surrounding the sequence to continue despite errors within the sequence, assuming they are recoverable. As we are only building the transformation script, it is preferable to propagate failures so we know when something did not apply.

To check or debug a transform sequence, it is possible to print various entities associated with the transform IR values. For example, we can print the operations associated with the handles:

\begin{mlirblock}
transform.sequence failures(propagate) {
^bb0(%arg0: !transform.any_op,
     %arg1: !transform.op<"linalg.matmul">,
     %arg2: !transform.op<"linalg.elemwise_binary">):
  transform.debug.emit_remark_at %arg1, "matmul"
      : !transform.op<"linalg.matmul">
  transform.debug.emit_remark_at %arg2, "elemwise_binaries"
      : !transform.op<"linalg.elemwise_binary">
  transform.yield
}
\end{mlirblock}

\subsection{Transform Dialect Interpreter}
Since we don’t want to recompile the compiler every time we change a transformation, we can use a Transform dialect interpreter pass to apply this transformation sequence to the payload IR. As we will see in the next chapter, it is possible to define custom passes or even integrate the transform interpreter into a larger pass. For now, we can use the existing test pass:

\begin{shellblock}
$ mlir-opt sequence.mlir --pass-pipeline="
    builtin.module(transform-interpreter{
        debug-bind-trailing-args=linalg.matmul,linalg.elemwise_binary})"
The sequence.mlir file contains both the payload IR function and the transform IR sequence nested in the same module. The transform interpreter pass will apply the @__transform_main named sequence to the anchor operation of the pass. In our case, we also asked the interpreter pass to associate the two extra arguments of the top-level sequence with all linalg.matmul and linalg.elemwise_binary payload operations through the respective pass options. Running this pass results in the expected remarks:

sequence.mlir:7:13: remark: matmul
  %matmul = linalg.matmul ins(%lhs, %rhs: tensor<512x512xf32>, tensor<512x512xf32>)
            ^
sequence.mlir:7:13: note: see current operation: %0 = linalg.matmul ins(%arg0, %arg1 : tensor<512x512xf32>, tensor<512x512xf32>) outs(%arg3 : tensor<512x512xf32>) -> tensor<512x512xf32>
sequence.mlir:10:13: remark: elemwise_binaries
  %biased = linalg.elemwise_binary { fun = #linalg.binary_fn<add> }
            ^
sequence.mlir:10:13: note: see current operation: %1 = linalg.elemwise_binary {fun = #linalg.binary_fn<add>} ins(%0, %arg2 : tensor<512x512xf32>, tensor<512x512xf32>) outs(%arg3 : tensor<512x512xf32>) -> tensor<512x512xf32>
sequence.mlir:14:13: remark: elemwise_binaries
  %relued = linalg.elemwise_binary { fun = #linalg.binary_fn<max_signed> }
            ^
sequence.mlir:14:13: note: see current operation: %2 = linalg.elemwise_binary {fun = #linalg.binary_fn<max_signed>} ins(%1, %cst : tensor<512x512xf32>, f32) outs(%arg3 : tensor<512x512xf32>) -> tensor<512x512xf32>
\end{shellblock}

Note that \mlircode{

\subsection{Specifying Transformations}

Now that we have handles to the operations we want to transform, we are ready to apply the transformations. Let us first try tiling the matmul operation itself.

\begin{mlirblock}
module attributes {transform.with_named_sequence} {
  transform.named_sequence @__transform_main(
       %arg0: !transform.any_op,
       %arg1: !transform.op<"linalg.matmul">,
       %arg2: !transform.op<"linalg.elemwise_binary">) {
    // The actual tiling transformation takes tile sizes as attributes.
    %loop, %tiled = transform.structured.tile_using_forall %arg1
                    tile_sizes [4, 32]
      : (!transform.op<"linalg.matmul">)
     -> (!transform.any_op, !transform.any_op)
    transform.yield
  }
}
\end{mlirblock}
The transformation returns two handles, as indicated in its documentation:

\begin{itemize}
\item A handle to \mlircode{linalg.generic} operating on the subset of the original data.
\item A handle to the \mlircode{scf.forall} “multi-for” loop around tensors.
\end{itemize}

Running this transformation with the same command as above expectedly produces the tiled code.

\begin{mlirblock}
func.func @fc_relu(%arg0: tensor<512x512xf32>,
                   %arg1: tensor<512x512xf32>,
                   %arg2: tensor<512x512xf32>,
                   %arg3: tensor<512x512xf32>) -> tensor<512x512xf32> {
  %cst = arith.constant 0.000000e+00 : f32
  %0 = scf.forall (%arg4, %arg5) in (128, 16) shared_outs(%arg6 = %arg3) -> (tensor<512x512xf32>) {
    %3 = affine.apply affine_map<(d0) -> (d0 * 4)>(%arg4)
    %4 = affine.apply affine_map<(d0) -> (d0 * 32)>(%arg5)
    %extracted_slice = tensor.extract_slice %arg0[%3, 0] [4, 512] [1, 1]
                     : tensor<512x512xf32> to tensor<4x512xf32>
    %extracted_slice_0 = tensor.extract_slice %arg1[0, %4] [512, 32] [1, 1]
                       : tensor<512x512xf32> to tensor<512x32xf32>
    %extracted_slice_1 = tensor.extract_slice %arg6[%3, %4] [4, 32] [1, 1]
                      : tensor<512x512xf32> to tensor<4x32xf32>
    %5 = linalg.matmul
         ins(%extracted_slice, %extracted_slice_0
             : tensor<4x512xf32>, tensor<512x32xf32>)
         outs(%extracted_slice_1 : tensor<4x32xf32>) -> tensor<4x32xf32>
    scf.forall.in_parallel {
      tensor.parallel_insert_slice %5 into %arg6[%3, %4] [4, 32] [1, 1]
          : tensor<4x32xf32> into tensor<512x512xf32>
    }
  }
  %1 = linalg.elemwise_binary {fun = #linalg.binary_fn<add>}
    ins(%0, %arg2 : tensor<512x512xf32>, tensor<512x512xf32>)
    outs(%arg3 : tensor<512x512xf32>) -> tensor<512x512xf32>
  %2 = linalg.elemwise_binary {fun = #linalg.binary_fn<max_signed>}
    ins(%1, %cst : tensor<512x512xf32>, f32)
    outs(%arg3 : tensor<512x512xf32>) -> tensor<512x512xf32>
  return %2 : tensor<512x512xf32>
}
\end{mlirblock}

Besides producing new handles, the tiling transform operation consumes the operand handle. This means that the handle is invalidated after this operation, and is no longer supposed to be used. Transform operations are required to mark all their operands as either consumed or readonly. Transform operations usually consume the operand if the associated payload operations are erased or recreated (which means erased and created anew with similar structure). As handles are essentially references to payload operations, they would become dangling if the payload no longer exists.

\subsection{Handle Invalidation and Expensive Checks Mode}

Undefined behavior is difficult to grapple with when it does happen, so the Transform dialect interpreter defaults to performing a set of additional, potentially expensive, checks that detect most undefined behavior in the transform IR. For example, if we wanted to use the \mlircode{

\begin{mlirblock}
module attributes {transform.with_named_sequence} {
  transform.named_sequence @__transform_main(
       %arg0: !transform.any_op,
       %arg1: !transform.op<"linalg.matmul">,
       %arg2: !transform.op<"linalg.elemwise_binary">) {
    // The actual tiling transformation takes tile sizes as attributes.
    %loop, %tiled = transform.structured.tile_using_forall %arg1 tile_sizes [4, 32]
        : (!transform.op<"linalg.matmul">) -> (!transform.any_op, !transform.any_op)

    // This is trying to use an invalidated handle leading to undefined behavior.
    transform.debug.emit_remark_at %arg1, "remark" : !transform.op<"linalg.matmul">
    transform.yield
  }
}
\end{mlirblock}

However, with the expensive checks enabled in the interpreter, a nice diagnostic is produced:

\begin{shellblock}
sequence.mlir:28:3: error: op uses a handle invalidated by a previously executed transform op
  transform.debug.emit_remark_at %mm, "elemwise_binaries" : !transform.any_op
  ^
sequence.mlir:26:9: note: handle to invalidated ops
  %mm = transform.cast %matmul : !transform.op<"linalg.matmul"> to !transform.any_op
        ^
sequence.mlir:27:19: note: invalidated by this transform op that consumes its operand #0 and invalidates all handles to payload IR entities associated with this operand and entities nested in them
  %loop, %tiled = transform.structured.tile_using_forall %mm tile_sizes [4, 32]
\end{shellblock}

When compile-time performance is a concern, and the transformation sequence is sufficiently stable, it is possible to disable expensive checks in the interpreter for improved performance by providing the disable-expensive-checks option to the pass or by setting the corresponding flag in the \cppcode{TransformOptions} passed into \cppcode{applyTransforms}.

One may observe that some operations such as transform.cast do not consume the operand (because they don’t erase the corresponding operation). So what would happen if we tried to use that operand instead?

\begin{mlirblock}
module attributes {transform.with_named_sequence} {
  transform.named_sequence @__transform_main
       %arg0: !transform.any_op,
       %arg1: !transform.op<"linalg.matmul">,
       %arg2: !transform.op<"linalg.elemwise_binary">) {
    // We can cast one type to another as long as operations are compatible
    // with both types. This creates "aliasing" handles.
    %casted = transform.cast %arg1 : !transform.op<"linalg.matmul">
        to !transform.any_op

    // The actual tiling transformation takes tile sizes as attributes.
    %loop, %tiled = transform.structured.tile_using_forall %arg1
                    tile_sizes [4, 32]
      : (!transform.op<"linalg.matmul">)
     -> (!transform.any_op, !transform.any_op)

    // Consuming an operand invalidates the consumed handle and any other handle
    // that is associated with the same payload operations, or payload
    // operations nested in them.
    transform.debug.emit_remark_at %casted, "remark"
      : !transform.any_op
    transform.yield
  }
}
\end{mlirblock}

Both \mlircode{

\begin{shellblock}
sequence.mlir:28:3: error: op uses a handle invalidated by a previously executed transform op
  transform.debug.emit_remark_at %matmul, "elemwise_binaries" : !transform.op<"linalg.matmul">
  ^
sequence.mlir:21:29: note: handle to invalidated ops
^bb0(%root: !transform.any_op, %matmul: !transform.op<"linalg.matmul">, %elemwise: !transform.op<"linalg.elemwise_binary">):
                            ^
sequence.mlir:27:19: note: invalidated by this transform op that consumes its operand #0 and invalidates all handles to payload IR entities associated with this operand and entities nested in them
  %loop, %tiled = transform.structured.tile_using_forall %mm tile_sizes [4, 32]
\end{shellblock}

\subsection{Chaining Transformations with Handles}

Going back to the transformation sequence, we have tiled the matrix multiplication, but we also want to tile and fuse the elementwise operations. The typical way of doing in the structured operations paradigm is to tile the last operation in some acyclic dataflow graph, and then progressively fuse the operations that produce its operands. This removes the need to explicitly tile all operations as fusion can adapt their sizes and inject recomputation if desired. So instead of tiling the matmul operation, we are going to tile the last operation in the chain, and then fuse the preceding operations into the loops produced by tiling.

\begin{mlirblock}
module attributes {transform.with_named_sequence} {
  transform.named_sequence @__transform_main(
       %arg0: !transform.any_op,
       %arg1: !transform.op<"linalg.matmul">,
       %arg2: !transform.op<"linalg.elemwise_binary">) {
    // Since the %arg2 handle is associated with both elementwise operations,
    // we need to split it into two handles so we can target only the second
    // elementwise operation.
    %add, %max = transform.split_handle %arg2
        : (!transform.op<"linalg.elemwise_binary">)
        -> (!transform.any_op, !transform.any_op)

    // The actual tiling transformation takes tile sizes as attributes. It
    // produces a handle to the loop generated during tiling.
    %tiled_max, %loop =
        transform.structured.tile_using_forall %max tile_sizes [8, 32]
          : (!transform.any_op) -> (!transform.any_op, !transform.any_op)

    // We can now fuse the other operations into the loop. Here, we fuse
    // operations one by one. This requires the operation that is being fused to
    // define the value used within the loop, so the order of such fusions is
    // important. We could also use "transform.merge_handles" to obtain a single
    // handle to all operations and give it to `fuse_into_containing_op` that
    // would take care of the ordering in this case.
    %add_fused, %loop_0 =
        transform.structured.fuse_into_containing_op %add into %loop
          : (!transform.any_op, !transform.any_op)
            -> (!transform.any_op, !transform.any_op)
    %matmul_fused, %loop_1 =
        transform.structured.fuse_into_containing_op %arg1 into %loop_0
          : (!transform.op<"linalg.matmul">, !transform.any_op)
            -> (!transform.any_op, !transform.any_op)

    transform.yield
  }
}
\end{mlirblock}

This achieves the desired tiling and fusion.

\subsection{More Handle Invalidation}
Finally, let us assume there exists an efficient microkernel, or a hardware instruction expressed as an intrinsic function, for a 4x4 matrix multiplication. For this purpose, we need to tile the fused operation to the desired size, and then outline it. The resulting function call can then be replaced with a call to the microkernel.

\begin{mlirblock}
module attributes {transform.with_named_sequence} {
  transform.named_sequence @__transform_main(
       %arg0: !transform.any_op,
       %arg1: !transform.op<"linalg.matmul">,
       %arg2: !transform.op<"linalg.elemwise_binary">) {
    // Since the %arg2 handle is associated with both elementwise operations,
    // we need to split it into two handles so we can target only the second
    // elementwise operation.
    %add, %max = transform.split_handle %arg2
        : (!transform.op<"linalg.elemwise_binary">)
          -> (!transform.any_op, !transform.any_op)

    // The actual tiling transformation takes tile sizes as attributes. It
    // produces a handle to the loop generated during tiling.
    %tiled, %loop = transform.structured.tile_using_forall %max
                    tile_sizes [8, 32]
        : (!transform.any_op) -> (!transform.any_op, !transform.any_op)

    // We can now fuse the other operations into the loop. Here, we fuse
    // operations one by one. This requires the operation that is being fused to
    // define the value used within the loop, so the order of such fusions is
    // important. We could also use "transform.merge_handles" to obtain a single
    // handle to all operations and give it to `fuse_into_containing_op` that
    // would take care of the ordering in this case.
    %add_fused, %loop_0 =
        transform.structured.fuse_into_containing_op %add into %loop
          : (!transform.any_op, !transform.any_op)
            -> (!transform.any_op, !transform.any_op)
    %matmul_fused, %loop_1 =
        transform.structured.fuse_into_containing_op %arg1 into %loop_0
          : (!transform.op<"linalg.matmul">, !transform.any_op)
            -> (!transform.any_op, !transform.any_op)

    // Tile again to get the desired size. Note that this time this tiles the
    // "add" operation and fuses matmul into the loop, but doesn't affect the
    // "max" operation. This illustrates the precise targeting with the
    // transform dialect. Otherwise, it is difficult to differentiate "add" and
    // "max", both of which having the same kind.
    %tiled_2, %loop_2 =
        transform.structured.tile_using_forall %add_fused tile_sizes [4, 4]
          : (!transform.any_op) -> (!transform.any_op, !transform.any_op)
    %matmul_fused_2, %loop_3 =
        transform.structured.fuse_into_containing_op %matmul_fused into %loop_2
          : (!transform.any_op, !transform.any_op)
            -> (!transform.any_op, !transform.any_op)

    // Since outlining is currently only implemented for region-holding
    // operations such as loops, use tiling to size 1 to materialize the outer
    // loop that is going to be outlined.
    %_, %outline_target =
        transform.structured.tile_using_forall %tiled_2 tile_sizes [1]
          : (!transform.any_op) -> (!transform.any_op, !transform.any_op)
    transform.structured.fuse_into_containing_op %matmul_fused_2
        into %outline_target
          : (!transform.any_op, !transform.any_op)
            -> (!transform.any_op, !transform.any_op)
    %func, %call = transform.loop.outline %outline_target
                   {func_name = "outlined"}
        : (!transform.any_op) -> (!transform.any_op, !transform.op<"func.call">)

    transform.yield
  }
}
\end{mlirblock}

This additional transformation also illustrates handle invalidation for nested operations. The transform.loop.outline operation consumes the handle to the loop, which invalidates it and all handles to any operations nested in it, such as \mlircode{

Attempting to access the fusion result after outlining produces the following error

\begin{shellblock}
test/Examples/transform/Ch1/invalidation-2.mlir:109:3: error: op uses a handle invalidated by a previously executed transform op
  transform.debug.emit_remark_at %outline_target, "outlined loop" : !transform.any_op
  ^
test/Examples/transform/Ch1/invalidation-2.mlir:102:25: note: handle to invalidated ops
  %outline_target, %_ = transform.structured.tile_using_forall %tiled_2 tile_sizes [1]
                        ^
test/Examples/transform/Ch1/invalidation-2.mlir:106:18: note: invalidated by this transform op that consumes its operand #0 and invalidates all handles to payload IR entities associated with this operand and entities nested in them
  %func, %call = transform.loop.outline %outline_target {func_name = "outlined"}
                 ^
test/Examples/transform/Ch1/invalidation-2.mlir:24:13: note: ancestor payload op
  %biased = linalg.elemwise_binary { fun = #linalg.binary_fn<add> }
            ^
test/Examples/transform/Ch1/invalidation-2.mlir:24:13: note: nested payload op
  %matmul = linalg.matmul ins(%lhs, %rhs: tensor<512x512xf32>, tensor<512x512xf32>)
\end{shellblock}

Note that the “add” elementwise operation is indicated as payload ancestor because it was used to produce the tile loop, and the loop therefore has its location.

Finally, we would like to replace the call to the outlined function with a call to the microkernel. Unfortunately, the Transform dialect doesn’t have support for this transformation (and cannot have if the call is rewritten to a custom, out-of-tree operation). Therefore, we need to define new transform operations. The next chapters will describe how this can be done.

\subsection{Tracking IR Modifications}

The Transform dialect automatically tracks all IR changes that are made as part of transform ops. (Implementations must use the provided rewriter to modify IR.) If a payload op is erased, it is automatically removed from all handles that it is currently associated with. If a payload op is replaced, the transform dialect tries to find the replacement op and updates all handles accordingly. If a multi-result op is replaced with values that are defined by multiple ops, or if an op is replaced with an op of a different type, an error is produced. This is because it is unclear whether the direct replacements actually represent the computation of the original op. There are ways to customize this behavior. More details can be found at the documentation of \cppcode{transform::TrackingListener}.

\section{Adding a Simple New Transformation Operation}

\subsection{Setting Up to Add New Transformations}

Before defining a new transform operation, we need to choose where its implementation should be located. While MLIR encourages upstream contributions, it is not always possible or even desirable to modify the main Transform dialect, for example, if the transformation is specific to some out-of-tree dialect that is not itself available upstream.

The Transform dialect uses the dialect extension mechanism to allow additional operations to be injected without modifying the dialect itself. Dialect extensions are registered with the context and loaded when the dialect itself is loaded. Extension definition is straightforward:

\begin{cppblock}
// In MyExtension.cpp.
#include "mlir/Dialect/Transform/IR/TransformDialect.h"

// Define a new Transform dialect extension. This uses the CRTP idiom to
// identify extensions.
class MyExtension : public ::mlir::transform::TransformDialectExtension<MyExtension> {
public:
  // The extension must derive the base constructor.
  using Base::Base;

  // This function initializes the extension, similarly to `initialize` in
  // dialect  definitions. List individual operations and dependent dialects
  // here.
  void init();
};

void MyExtension::init() {
  // Similarly to dialects, an extension can declare a dependent dialect. This
  // dialect will be loaded along with the extension and, therefore, along with
  // the Transform  dialect. Only declare as dependent the dialects that contain
  // the attributes or types used by transform operations. Do NOT declare as
  // dependent the dialects produced during the transformation.
  //
  // declareDependentDialect<MyDialect>();

  // When transformations are applied, they may produce new operations from
  // previously unloaded dialects. Typically, a pass would need to declare
  // itself dependent on the dialects containing such new operations. To avoid
  // confusion with the dialects the extension itself depends on, the Transform
  // dialects differentiates between:
  //   - dependent dialects, which are used by the transform operations, and
  //   - generated dialects, which contain the entities (attributes, operations,
  //     types) that may be produced by applying the transformation even when
  //     not present in the original payload IR.
  // In the following chapter, we will be add operations that generate function
  // calls and structured control flow operations, so let's declare the
  // corresponding dialects as generated.
  declareGeneratedDialect<::mlir::scf::SCFDialect>();
  declareGeneratedDialect<::mlir::func::FuncDialect>();

  // Finally, we register the additional transform operations with the dialect.
  registerTransformOps<
    // TODO: list the operation classes.
  >();
}
\end{cppblock}

The operations themselves can be defined using ODS, exactly in the same way as regular operations in a dialect.

\begin{tdblock}
// In MyExtension.td
#ifndef MY_EXTENSION
#define MY_EXTENSION

include "mlir/Dialect/Transform/IR/TransformDialect.td"
include "mlir/Dialect/Transform/Interfaces/TransformInterfaces.td"
include "mlir/IR/OpBase.td"
include "mlir/Interfaces/SideEffectInterfaces.td"

def MyOp : Op<Transform_Dialect, "transform.my.op", [
    // TODO: interfaces and traits here.
   ]> {
  let summary = "my transform op";
  // TODO: define the operation properties.
}

#endif // MY_EXTENSION
\end{tdblock}

Similarly to dialects, we must use Tablegen to generate the header and implementation of these operations. We can instruct CMake to do it as follows.

\begin{cmakeblock}
# In CMakeLists.txt next to MyExtension.td.

# Tell Tablegen to use MyExtension.td as input.
set(LLVM_TARGET_DEFINITIONS MyExtension.td)

# Ask Tablegen to generate op declarations and definitions from ODS.
mlir_tablegen(MyExtension.h.inc -gen-op-decls)
mlir_tablegen(MyExtension.cpp.inc -gen-op-defs)

# Add a CMakeTarget we can depend on to ensure the generation happens before the compilation.
add_public_tablegen_target(MyExtensionIncGen)

# Don't forget to generate the documentation, this will produce a MyExtension.md under
# Dialects.
add_mlir_doc(MyExtension MyExtension Dialects/ -gen-op-doc)
# In CMakeLists.txt next to MyExtension.cpp
add_mlir_library(
  # Library called MyExtension.
  MyExtension

  # Built from the following source files.
  MyExtension.cpp

  # Make sure ODS declaration and definitions are generated before compiling
  # this.
  DEPENDS
  MyExtensionIncGen

  # Link in the transform dialect, and all generated dialects.
  LINK_LIBS PUBLIC
  MLIRTransformDialect
  MLIRFuncDialect
  MLIRSCFDialect
)
\end{cmakeblock}

This will generate two files, \cppcode{MyExtension.h.inc} and \cppcode{MyExtension.cpp.inc}, that are supposed to be included into the declaration and definition of the transform operations, respectively.

\begin{cppblock}
// In MyExtension.h.
#include "mlir/Dialect/Transform/IR/TransformDialect.h"
#include "mlir/Dialect/Transform/Interfaces/TransformInterfaces.h"

#define GET_OP_CLASSES
#include "MyExtension.h.inc"
// In MyExtension.cpp.

#define GET_OP_CLASSES
#include "MyExtension.cpp.inc"

// …
void MyExtension::init() {
  // …

  // Finally, we register the additional transform operations with the dialect.
  // List all  operations generated from ODS. This call will perform additional
  // checks that the  operations implement the transform and memory effect
  // interfaces required by the dialect interpreter and assert if they do not.
  registerTransformOps<
#define GET_OP_LIST
#include "MyExtension.cpp.inc"
  >();
}
\end{cppblock}

\subsection{Defining a Transform Operation}

With this setup, we are now ready to define the new transform operation to rewrite the function call. This is identical to defining a regular operation in a dialect. Note that the Transform dialect requires operations to implement the \cppcode{TransformOpInterface} as well as \cppcode{MemoryEffectsOpInterface} to indicate whether the operands are consumed or only read. Our operation can be defined along the following lines.

\begin{tdblock}
// In MyExtension.td.

// Define the new operation. By convention, prefix its name with the name of the
// dialect  extension, "my.". The full operation name will be further prefixed
// with "transform.".
def ChangeCallTargetOp : Op<Transform_Dialect, "my.change_call_target",
    // Indicate that the operation implements the required TransformOpInterface
    // and MemoryEffectsOpInterface.
    [DeclareOpInterfaceMethods<TransformOpInterface>,
     DeclareOpInterfaceMethods<MemoryEffectsOpInterface>]> {
  // Provide a brief and a full description. It is recommended that the latter
  // describes the effects on the operands and how the operation processes
  // various failure modes.
  let summary = "Changes the callee of a call operation to the specified one";
  let description = [{
    For each `func.call` payload operation associated with the handle, changes
    its callee to be the symbol whose name is provided as an attribute to this operation.

    Generates a silenceable failure if the operand is associated with payload operations that are not `func.call`. Only reads the operand.
  }];

  // The arguments include the handle to the payload operations and the
  // attribute that specifies the new callee. The handle must implement
  // TransformHandleTypeInterface.
  // We use a string attribute as the symbol may not exist in the transform IR
  // so the verification may fail.
  let arguments = (ins
    TransformHandleTypeInterface:$call,
    StrAttr:$new_target);

  // The results are empty as the transformation does not produce any new
  // payload.
  let results = (outs);

  // Provide nice syntax.
  let assemblyFormat = "$call `,` $new_target attr-dict `:` type($call)";
}
\end{tdblock}

To finalize the definition of the transform operation, we need to implement the interface methods. The \cppcode{TransformOpInterface} currently requires only one method---\cppcode{apply}---that performs the actual transformation. It is a good practice to limit the body of the method to manipulation of the Transform dialect constructs and have the actual transformation implemented as a standalone function so it can be used from other places in the code. Similar to rewrite patterns, all IR must be modified with the provided rewriter.

\begin{cppblock}
// In MyExtension.cpp

// Implementation of our Transform dialect operation.
// This operation returns a tri-state result that can be one of:
// - success when the transformation succeeded;
// - definite failure when the transformation failed in such a way that
//   following transformations are impossible or undesirable, typically it could
//   have left payload IR in an invalid state; it is expected that a diagnostic
//   is emitted immediately before returning the definite error;
// - silenceable failure when the transformation failed but following
//   transformations are still applicable, typically this means a precondition
//   for the transformation is not satisfied and the payload IR has not been
//   modified. The silenceable failure additionally carries a Diagnostic that
//   can be emitted to the user.
::mlir::DiagnosedSilenceableFailure mlir::transform::ChangeCallTargetOp::apply(
    // The rewriter that should be used when modifying IR.
    ::mlir::transform::TransformRewriter &rewriter,
    // The list of payload IR entities that will be associated with the
    // transform IR values defined by this transform operation. In this case, it
    // can remain empty as there are no results.
    ::mlir::transform::TransformResults &results,
    // The transform application state. This object can be used to query the
    // current associations between transform IR values and payload IR entities.
    // It can also carry additional user-defined state.
    ::mlir::transform::TransformState &state) {

  // First, we need to obtain the list of payload operations that are associated
  // with the operand handle.
  auto payload = state.getPayloadOps(getCall());

  // Then, we iterate over the list of operands and call the actual IR-mutating
  // function. We also check the preconditions here.
  for (Operation *payloadOp : payload) {
    auto call = dyn_cast<::mlir::func::CallOp>(payloadOp);
    if (!call) {
      DiagnosedSilenceableFailure diag = emitSilenceableError()
          << "only applies to func.call payloads";
      diag.attachNote(payloadOp->getLoc()) << "offending payload";
      return diag;
    }

    updateCallee(call, getNewTarget());
  }

  // If everything went well, return success.
  return DiagnosedSilenceableFailure::success();
}
\end{cppblock}

The implementation of the \cppcode{MemoryEffectsOpInterface} must specify the effects this operation has on its operands (consumed or readonly) and on the payload IR (mutates or readonly). Transform dialect verifiers will check for side effects being present and assert in debug builds if they are not.

\begin{cppblock}
// In MyExtension.cpp

void ChangeCallTargetOp::getEffects(
    ::llvm::SmallVectorImpl<::mlir::MemoryEffects::EffectInstance> &effects) {
  // Indicate that the `call` handle is only read by this operation because the
  // associated operation is not erased but rather modified in-place, so the
  // reference to it remains valid.
  onlyReadsHandle(getCall(), effects);

  // Indicate that the payload is modified by this operation.
  modifiesPayload(effects);
}
\end{cppblock}

\subsection{Registration and Usage}
This is enough to define transform operations. The only remaining bit is providing the extension registration hook that can be called from the project’s main.

\begin{cppblock}
// In TransformDialect.cpp (don't forget a declaration in TransformDialect.h);

void registerMyExtension(::mlir::DialectRegistry &registry) {
  registry.addExtensions<MyExtension>();
}
\end{cppblock}

After registering the extension, it becomes possible to use our new operation in the Transform dialect interpreter. The upstream testing pass can be used as is.

\begin{mlirblock}
module attributes {transform.with_named_sequence} {
  transform.named_sequence @__transform_main(
      %arg0: !transform.any_op,
      %arg1: !transform.op<"linalg.matmul">,
      %arg2: !transform.op<"linalg.elemwise_binary">) {
    // Since the %arg2 handle is associated with both elementwise operations,
    // we need to split it into two handles so we can target only the second
    // elementwise operation.
    %add, %max = transform.split_handle %arg2
        : (!transform.op<"linalg.elemwise_binary">)
        -> (!transform.any_op, !transform.any_op)

    // The actual tiling transformation takes tile sizes as attributes. It
    // produces a handle to the loop generated during tiling.
    %loop, %tiled = transform.structured.tile_using_forall %max
                    tile_sizes [8, 32]
        : (!transform.any_op) -> (!transform.any_op, !transform.any_op)

    // We can now fuse the other operations into the loop. Here, we fuse
    // operations one-by-one. This requires the operation that is being fused
    // to define the value used within the loop, so the order of such fusions
    // is important. We could also use "transform.merge_handles" to obtain
    // a single handle to all operations and give it to
    // `fuse_into_containing_op` that would take care of the ordering in this
    // case.
    %add_fused = transform.structured.fuse_into_containing_op %add into %loop
        : (!transform.any_op, !transform.any_op) -> !transform.any_op
    %matmul_fused = transform.structured.fuse_into_containing_op %arg1
                    into %loop
        : (!transform.op<"linalg.matmul">, !transform.any_op)
       -> !transform.any_op

    // Tile again to get the desired size. Note that this time this tiles the
    // "add" operation and fuses matmul into the loop, but doesn't affect the
    // "max" operation. This illustrates the precise targeting with the
    // transform dialect. Otherwise, it is difficult to differentiate "add" and
    // "max", both of which having the same kind.
    %loop_2, %tiled_2 = transform.structured.tile_using_forall %add_fused
                        tile_sizes [4, 4]
        : (!transform.any_op) -> (!transform.any_op, !transform.any_op)
    %matmul_fused_2 = transform.structured.fuse_into_containing_op %matmul_fused
                      into %loop_2
        : (!transform.any_op, !transform.any_op) -> !transform.any_op

    // Since outlining is currently only implemented for region-holding
    // operations such as loops, use tiling to size 1 to materialize the outer
    // loop that is going to be outlined.
    %outline_target, %_ = transform.structured.tile_using_forall %tiled_2 tile_sizes [1]
        : (!transform.any_op) -> (!transform.any_op, !transform.any_op)
    transform.structured.fuse_into_containing_op %matmul_fused_2 into %outline_target
        : (!transform.any_op, !transform.any_op) -> !transform.any_op
    %func, %call = transform.loop.outline %outline_target
                   {func_name = "outlined"}
        : (!transform.any_op) -> (!transform.any_op, !transform.any_op)

    // Rewrite the call target.
    transform.my.change_call_target %call, "microkernel" : !transform.any_op

    transform.yield
  }
}
\end{mlirblock}

\section{More than Simple Transform Operations}
\subsection{Type Constraints and {\tt ApplyEach} Trait}
A transform operation that applies to each payload operation individually and requires it to be of a specific kind is a repeated pattern. One can use Transform dialect types to specify the preconditions of the type. Specifically, we can change the expected operand type from the wide \cppcode{TransformHandleTypeInterface} to the more narrow \cppcode{Transform_ConcreteOp<"func.call">}. Furthermore, we use the \cppcode{TransformEachOpTrait} trait to provide the skeleton implementation of the apply method that performs verification, iteration over payloads and result concatenation. The improved ODS op definition is as follows.

\begin{tdblock}
// In MyExtension.td.

// Define the new operation. By convention, prefix its name with the name of the 
// dialect extension, "my.". The full operation name will be further prefixed with 
// "transform.".
def ChangeCallTargetOp : Op<Transform_Dialect, "my.change_call_target",
    // Indicate that the operation implements the required TransformOpInterface and
    // MemoryEffectsOpInterface. Use the TransformEach trait to provide the
    // implementation for TransformOpInterface.
    [TransformOpInterface, TransformEachOpTrait,
     DeclareOpInterfaceMethods<MemoryEffectsOpInterface>]> {
  // Provide a brief and a full description. It is recommended that the latter describes
  // the effects on the operands and how the operation processes various failure modes.
  let summary = "Changes the callee of a call operation to the specified one";
  let description = [{
    For each `func.call` payload operation associated with the handle, changes its
    callee to be the symbol whose name is provided as an attribute to this operation.

    Generates a silenceable failure if the operand is associated with payload operations
    that are not `func.call`.
    Only reads the operand.
  }];

  // The arguments include the handle to the payload operations and the attribute that
  // specifies the new callee. The handle must implement TransformHandleTypeInterface.
  // We use a string attribute as the symbol may not exist in the transform IR so the
  // verification may fail.
  let arguments = (ins
    Transform_ConcreteOpType<"func.call">:$call,
    StrAttr:$new_target);

  // The results are empty as the transformation does not produce any new payload.
  let results = (outs);

  // Provide nice syntax.
  let assemblyFormat = "$call `,` $new_target attr-dict `:` type($call)";

  // Declare the function implementing the interface for a single payload operation.
  let extraClassDeclaration = [{
    ::mlir::DiagnosedSilenceableFailure applyToOne(
        ::mlir::transform::TransformRewriter &rewriter,
        ::mlir::func::CallOp call,
        ::mlir::transform::ApplyToEachResultList &results,
        ::mlir::transform::TransformState &state);
  }];
}
\end{tdblock}

Now, instead of defining the apply method with a loop, we can simply define a function that applies to an individual payload operation and the trait will take care of the rest.

\begin{cppblock}
::mlir::DiagnosedSilenceableFailure ChangeCallTargetOp::applyToOne(
    ::mlir::transform::TransformRewriter &rewriter,
    ::mlir::func::CallOp call,
    ::mlir::transform::ApplyToEachResultList &results,
    ::mlir::transform::TransformState &state) {
  // Call the actual transformation function.
  updateCallee(call, getNewTarget());
  // Indicate success.
  return DiagnosedSilenceableFailure::success();
}
\end{cppblock}

\subsection{Defining a Transform Type}

In addition to operations, the Transform dialect allows extensions to define and inject additional attributes and types. As we have seen above, transform types are used to specify constraints on the payload operations. Our call rewriting operation currently applies only to \mlircode{func.call}. We may want to generalize it to apply to any payload operation that implements \cppcode{CallOpInterface}, but the Transform dialect currently doesn’t provide a type that checks if a payload operation implements this interface. Let’s define it in our extension.

Type definition is again identical to defining a dialect type with ODS.

\begin{tdblock}
// Transform dialect allows additional types to be defined and injected.
def CallOpInterfaceHandle
  : TypeDef<Transform_Dialect, "CallOpInterfaceHandle",
      // The type must implement `TransformHandleTypeInterface`.
      [DeclareTypeInterfaceMethods<TransformHandleTypeInterface>]> {

  // The usual components of a type such as description, mnemonic and assembly format
  // should be provided.
  let summary = "handle to payload operations implementing CallOpInterface";
  let mnemonic = "my.call_op_interface";
  let assemblyFormat = "";
}
\end{tdblock}

We will omit the generation of declaration and definitions using Tablegen for brevity as it is identical to the regular case.

To finalize the definition of a transform type, one must implement the interface methods.

\begin{cppblock}
// In MyExtension.cpp.

// The interface declares this method to verify constraints this type has on
// payload operations. It returns the now familiar tri-state result.
mlir::DiagnosedSilenceableFailure
mlir::transform::CallOpInterfaceHandleType::checkPayload(
    // Location at which diagnostics should be emitted.
    mlir::Location loc,
    // List of payload operations that are about to be associated with the
    // handle that has this type.
    llvm::ArrayRef<mlir::Operation *> payload) const {

  // All payload operations are expected to implement CallOpInterface, check this.
  for (Operation *op : payload) {
    if (llvm::isa<mlir::CallOpInterface>(op))
      continue;

    // By convention, these verifiers always emit a silenceable failure since they are
    // checking a precondition.
    DiagnosedSilenceableFailure diag = emitSilenceableError(loc)
        << "expected the payload operation to implement CallOpInterface";
    diag.attachNote(op->getLoc()) << "offending operation";
    return diag;
  }

  // If everything is okay, return success.
  return DiagnosedSilenceableFailure::success();
}
\end{cppblock}

Additional attributes and types need to be registered in the extension, next to operations.

\begin{cppblock}
// In MyExtension.cpp.

void MyExtension::init() {
  // ...

  registerTypes<
#define GET_TYPEDEF_LIST
#include "MyExtensionTypes.cpp.inc"
  >();
}
\end{cppblock}

This type is now directly available in the Transform dialect and can be used in operations.

\begin{mlirblock}
// Cast to our new type.
%casted = transform.cast %call : !transform.any_op to !transform.my.call_op_interface
// Using our new operation.
transform.my.change_call_target %casted, "microkernel" : !transform.my.call_op_interface
\end{mlirblock}
 
\subsection{Operand Consumption}

As an exercise, let us modify the rewriting operation to consume the operand. This would be necessary, for example, if the transformation were to rewrite the \mlircode{func.call} operation into a custom operation my.mm4. Since the operand handle is now consumed, the operation can return a new handle to the newly produced payload operation. Otherwise, the ODS definition of the transform operation remains unchanged.

\begin{tdblock}
// In MyExtension.td.

// Define another transform operation.
def CallToOp : Op<Transform_Dialect, "my.call_to_op",
     // Indicate that the operation implements the required TransformOpInterface and
     // MemoryEffectsOpInterface. Use the TransformEach trait to provide the
     // implementation for TransformOpInterface.
    [TransformOpInterface, TransformEachOpTrait,
     DeclareOpInterfaceMethods<MemoryEffectsOpInterface>]> {
  // Summary and description omitted for brevity.

  // The argument is the handle to the payload operations.
  let arguments = (ins CallOpInterfaceHandle:$call);

  // The result is the handle to the payload operations produced during the
  // transformation.
  let results = (outs TransformHandleTypeInterface:$transformed);

  // Provide nice syntax.
  let assemblyFormat = "$call attr-dict `:` functional-type(inputs, outputs)";

  // Declare the function implementing the interface for a single payload operation.
  let extraClassDeclaration = [{
    ::mlir::DiagnosedSilenceableFailure applyToOne(
        ::mlir::transform::TransformRewriter &rewriter,
        ::mlir::CallOpInterface call,
        ::mlir::transform::ApplyToEachResultList &results,
        ::mlir::transform::TransformState &state);
  }];
}
\end{tdblock}

Now let’s look at the implementation of interface methods.

\begin{cppblock}
// In MyExtension.cpp.

::mlir::DiagnosedSilenceableFailure CallToOp::applyToOne(
    ::mlir::transform::TransformRewriter &rewriter,
    ::mlir::CallOpInterface call,
    ::mlir::transform::ApplyToEachResultList &results,
    ::mlir::transform::TransformState &state) {
  // Call the actual rewrite.
  Operation *rewritten = rewriteToOp(call);

  // Report an error if the rewriter produced a null pointer. Note that it may have
  // irreversibly modified the payload IR, so we produce a definite failure.
  if (!rewritten) {
    return emitDefiniteError() << "failed to rewrite call to operation";
  }

  // On success, push the resulting operation into the result list. The list is expected
  // to contain exactly one entity per result and per application. The handles will be
  // associated with lists of the respective values produced by each application.
  results.push_back(rewritten);

  // If everything is fine, return success.
  return DiagnosedSilenceableFailure::success();
}

void CallToOp::getEffects(
    ::llvm::SmallVectorImpl<::mlir::MemoryEffects::EffectInstance> &effects) {
  // Indicate using side effects that the operand handle is consumed, and the
  // result handle is produced.
  consumesHandle(getCall(), effects);
  producesHandle(getTransformed(), effects);

  // Indicate that the payload IR is modified.
  modifiesPayload(effects);
}
\end{cppblock}

The overall flow of these implementations is similar to the previous one. The application also needs to specify the resulting entities that are going to be associated with the handles it produces. Operations are required to specify the entities to associate with all results on success, even if the list is empty. An assertion will be triggered if it is not the case. In case of failure, the interpreter will automatically associate all results that are not yet defined with empty lists.

Note that since \cppcode{applyToOne} always expects one payload entity to be associated with each result handle in each application, it cannot be used to return handles associated with empty lists for non-empty operand handles. Instead, one would use apply directly.

\begin{cppblock}
::mlir::DiagnosedSilenceableFailure SomeOtherOp::apply(
    ::mlir::transform::TransformRewriter &rewriter,
    ::mlir::transform::TransformResults &results,
    ::mlir::transform::TransformState &state) {
  // ...

  // Associate the result `transformed` with an empty list of payload operations.
  results.set(cast<OpResult>(getTransformed()), {});
  return DiagnosedSilenceableFailure::success();
}
\end{cppblock}

\subsection{Memory Effects Traits}

Some common memory effect patterns are also available as traits to minimize the boilerplate.

\cppcode{FunctionalStyleTransformOpTrait} indicates that all handle-typed operands are consumed, all results are produced, and the payload IR is modified.
\cppcode{NavigationTransformOpTrait} indicates that all handle-typed operands are only read, all results are produced, and the payload IR is only read.
Using these traits removes the need to declare or define the methods of the \cppcode{MemoryEffectsOpInterface}.

\begin{tdblock}
// In MyExtension.td.

// Define another transform operation.
def CallToOp : Op<Transform_Dialect, "my.call_to_op",
     // Indicate that the operation implements the required TransformOpInterface.
     // Use the TransformEach trait to provide implementation of this interface.
    [TransformOpInterface, TransformEachOpTrait,
     // Indicate that the operation implements the required MemoryEffectsOpInterface.
     // Use the FunctionalStyle trait to provide the implementation for this interface.
     MemoryEffectsOpInterface, FunctionalStyleTransformOpTrait]> {
  // Summary and description omitted for brevity.

  // The argument is the handle to the payload operations.
  let arguments = (ins CallOpInterfaceHandle:$call);

  // The result is the handle to the payload operations produced during the
  // transformation.
  let results = (outs TransformHandleTypeInterface:$transformed);

  // Provide nice syntax.
  let assemblyFormat = "$call attr-dict `:` functional-type(operands, results)";

  // Declare the function implementing the interface for a single payload operation.
  let extraClassDeclaration = [{
    ::mlir::DiagnosedSilenceableFailure applyToOne(
        ::mlir::transform::TransformRewriter &rewriter,
        ::mlir::CallOpInterface call,
        ::mlir::transform::ApplyToEachResultList &results,
        ::mlir::transform::TransformState &state);
  }];
}
\end{tdblock}

\section{Matching Payload with Transform Operations}

Up until now, we were applying transform dialect scripts under the assumption that specific payload operations are identified by the caller when the transform dialect interpreter is invoked. This may be seen as contrary to the idea of driving transformations from a dialect since the transformation targets must be identified through mechanisms external to the transform dialect interpreter, for example, when invoking the interpreter programmatically in C++ or through pass arguments as seen in previous chapters. It also adds practical overhead due to increased interaction with the interpreter in C++, and cognitive overhead of manipulating two interfaces at once. To remedy this, Transform dialect proposes a subset of operations for matching payload operations that need to be transformed.

Match operations are simply transform operations with some additional guarantees. In particular, they are not expected to modify the payload IR and are expected to fail if their operands (typically payload operation handles) are not associated with payload IR objects having desired properties, such as operation names or kinds of arguments. Using simple combinator operations, it becomes possible to set up a higher-level match and rewrite infrastructure directly within the transform dialect.

\subsection{Simple Match}

Let us reconsider the ``fully connected layer'' example from Section~2, reproduced below for convenience.

\begin{mlirblock}
// Original function to optimize.
func.func @fc_relu(%lhs: tensor<512x512xf32>, %rhs: tensor<512x512xf32>,
                   %bias: tensor<512x512xf32>, %output: tensor<512x512xf32>)
                   -> tensor<512x512xf32> {
  // Matrix-matrix multiplication.
  %matmul = linalg.matmul
            ins(%lhs, %rhs: tensor<512x512xf32>, tensor<512x512xf32>)
            outs(%output: tensor<512x512xf32>) -> tensor<512x512xf32>

  // Elementwise addition.
  %biased = linalg.elemwise_binary { fun = #linalg.binary_fn<add> }
    ins(%matmul, %bias : tensor<512x512xf32>, tensor<512x512xf32>)
    outs(%output : tensor<512x512xf32>) -> tensor<512x512xf32>

  // Elementwise max with 0 (ReLU).
  %c0f = arith.constant 0.0 : f32
  %relued = linalg.elemwise_binary { fun = #linalg.binary_fn<max_signed> }
    ins(%biased, %c0f : tensor<512x512xf32>, f32)
    outs(%output : tensor<512x512xf32>) -> tensor<512x512xf32>
  func.return %relued : tensor<512x512xf32>
}
\end{mlirblock}

In Section~2, we were calling the test transform interpreter pass with additional arguments, \mlircode{bind-first-extra-to-ops=linalg.matmul bind-second-extra-to-ops=linalg.elemwise_binary}, to provide initial associations for operation handles. Instead, we can use match operations to discover relevant operations in the payload IR. Match operations can be combined with ``regular'' transform operations using, e.g., the \mlircode{transform.collect_matching} combinator operation that leverages the concept of named sequences to organize matchers.

\begin{mlirblock}
// The module containing named sequences must have an attribute allowing them
// to enable verification.
module @transforms attributes { transform.with_named_sequence } {
  // Entry point. This takes as the only argument the root operation (typically
  // pass root) given to the transform interpreter.
  transform.named_sequence @__transform_main(
      %root: !transform.any_op {transform.readonly}) {
    // Collect operations that match the criteria specified in named sequence.
    // If the named sequence fails with a silenceable failure, silences it (the
    // message is forwarded to the debug stream). If the named sequence
    // succeeds, appends its results to the results of this operation.
    %elemwise = transform.collect_matching @match_elemwise in %root
      : (!transform.any_op) -> !transform.any_op
    %matmul = transform.collect_matching @match_matmul in %root
      : (!transform.any_op) -> !transform.any_op
    transform.include @print_elemwise failures(propagate)  (%elemwise)
      : (!transform.any_op) -> ()
    transform.include @print_matmul failures(propagate)  (%matmul)
      : (!transform.any_op) -> ()

    transform.yield
  }

  // This is a matcher sequence. It is given an operation to match and the
  // match is considered successful unless any nested operation produces a
  // failure. The values yielded by this operation will be forwarded to the
  // rewriter sequence on success.
  transform.named_sequence @match_elemwise(
      %entry: !transform.any_op {transform.readonly}) -> !transform.any_op {
    transform.match.operation_name %entry ["linalg.elemwise_binary"]
      : !transform.any_op
    transform.yield %entry : !transform.any_op
  }
  transform.named_sequence @match_matmul(
      %entry: !transform.any_op {transform.readonly}) -> !transform.any_op {
    transform.match.operation_name %entry ["linalg.matmul"] : !transform.any_op
    transform.yield %entry : !transform.any_op
  }

  // This is a rewriter sequence.
  transform.named_sequence @print_elemwise(
      %elemwise_binary: !transform.any_op {transform.readonly}) {
    transform.debug.emit_remark_at
      %elemwise_binary, "elementwise binary" : !transform.any_op
    transform.yield
  }
  transform.named_sequence @print_matmul(
      %matmul: !transform.any_op {transform.readonly}) {
    transform.debug.emit_remark_at %matmul, "matmul" : !transform.any_op
    transform.yield
  }
}
\end{mlirblock}

This script can be executed using the non-test interpreter pass running on the root operation of the translation unit without additional flags: \mintinline{shell}{mlir-opt --transform-interpreter}. It will emit corresponding remarks at \mlircode{linalg.elemwise_binary} and \mlircode{linalg.matmul} operations. In debug builds, the infrastructure provides a convenient method to understand the matching process by passing \mintinline{shell}{-debug-only=transform-matcher} to \mintinline{text}{mlir-opt} or a derived tool. It will print the silenceable failure messages produced by the match operations into the debug stream, for example:

\begin{shellblock}
<...>
[transform-matcher] matching %0 = linalg.matmul ins(%arg0, %arg1 : tensor<512x512xf32>, tensor<512x512xf32>) outs(%arg3 : tensor<512x512xf32>) -> tensor<512x512xf32> @0x5622eee08410
[transform-matcher] matcher match_elemwise failed: wrong operation name
<...>
\end{shellblock}
This is now sufficient to run the rest of the transform script from Chapter 1, substituting \mlircode{

\subsection{Matching Chains of Operations}
The matcher above remains naive as it matches all operations of the certain kind under the payload root. These operations may or may not be related, and may, for example, belong to different functions. Even if they are in a single function, if there are multiple groups of such operations, we wouldn’t be able to differentiate them with this approach. In reality, we want to match a specific group of operations where a matmul operation produces a result that is used by an elementwise operation, which in turn feeds another elementwise operation in a similar way.

This can be achieved using the following matcher sequence.

\begin{mlirblock}
// This is also a matcher sequence. It is similarly given an operation to
// match and nested operations must succeed in order for a match to be deemed
// successful. It starts matching from the last operation in the use-def chain
// and goes back because each operand (use) has exactly one definition.
transform.named_sequence @match_matmul_elemwise(
    %last: !transform.any_op {transform.readonly})
    -> (!transform.any_op, !transform.any_op, !transform.any_op) {
  // The last operation must be an elementwise binary.
  transform.match.operation_name %last ["linalg.elemwise_binary"]
    : !transform.any_op
  // Its first operand must be defined by another operation, to which we
  // will get a handle here. We are guaranteed that the first operand exists
  // because we know the operation is binary, but even in absence of such a
  // guarantee, this operation would have produced a silenceable failure when
  // `%last` does not have enough operands.
  %middle = transform.get_producer_of_operand %last[0]
    : (!transform.any_op) -> !transform.any_op
  // The defining operation must itself be an elementwise binary.
  transform.match.operation_name %middle ["linalg.elemwise_binary"]
    : !transform.any_op
  // And the first operand of that operation must be defined by yet another
  // operation.
  %matmul = transform.get_producer_of_operand %middle[0]
    : (!transform.any_op) -> !transform.any_op
  // And that operation is a matmul.
  transform.match.operation_name %matmul ["linalg.matmul"] : !transform.any_op
  // We will yield the handles to the matmul and the two elementwise
  // operations separately.
  transform.yield %matmul, %middle, %last
    : !transform.any_op, !transform.any_op, !transform.any_op
}
\end{mlirblock}

This matcher is applicable in presence of other elemwise and matmul operations and will return the triple of related operations rather than operations in the order in which they are found. It can be exercised similarly to the previous incarnation, as follows.

\begin{mlirblock}
// Alternative entry point.
transform.named_sequence @__transform_main(
    %root: !transform.any_op {transform.readonly}) {
  // Collect groups of operations that match the criteria specified in the
  // named sequence.
  %matmul, %el1, %el2 = transform.collect_matching @match_matmul_elemwise in %root
    : (!transform.any_op) -> (!transform.any_op, !transform.any_op, !transform.any_op)
  %elemwise = transform.merge_handles %el1, %el2 : !transform.any_op

  transform.include @print_elemwise failures(propagate)  (%elemwise)
    : (!transform.any_op) -> ()
  transform.include @print_matmul failures(propagate)  (%matmul)
    : (!transform.any_op) -> ()

  transform.yield
}
\end{mlirblock}

\subsection{Defining Match Operations}
The matcher of a chain of operations is correct in presence of other operations, but is still insufficiently robust for many cases of interest. In particular, using \mlircode{transform.get_producer_of_operand 

Match operations are defined similarly to other transform operations, with the only difference of additionally implementing the \cppcode{MatchOpInterface}. Note that this interface has no additional methods (though it may add some eventually) and is only used as a verification contract that the operation is intended for matching and will not attempt to transform the payload. The minimal definition of our operation is as follows.

\begin{tdblock}
// Define the new operation. By convention, prefix its name with `match`
// followed by the name of the dialect extension.
def HasOperandSatisfyingOp : TransformDialectOp<"match.my.has_operand_satisfying",
    [DeclareOpInterfaceMethods<MemoryEffectsOpInterface>,
     DeclareOpInterfaceMethods<TransformOpInterface>,
     // Indicate that the operation implements MatchOpInterface in addition to
     // the TransformOpInterface. This interface is only used as a tag at this
     // point and has no methods that are mandatory to implement.
     MatchOpInterface,
     SingleBlockImplicitTerminator<"::mlir::transform::YieldOp">]> {
  let summary = "Succeed if any of the operands matches all nested criteria";
  let arguments = (ins TransformHandleTypeInterface:$op);
  let results = (outs TransformParamTypeInterface:$position,
                      Variadic<Transform_AnyHandleOrParamType>:$results);

  // Match operations can be arbitrarily complex, e.g., containing regions.
  let regions = (region SizedRegion<1>:$body);
  let hasVerifier = 1;
  let assemblyFormat = [{
    $op `:` functional-type($op, results) attr-dict-with-keyword $body
  }];
}
\end{tdblock}

It takes as argument the handle associated with the payload operations whose operands it will match, has an associated single-block region containing the match criteria, and returns the position of the matched operand as well as any other transform value yielded from the body on the successful match.

The matching logic is implemented in the apply method of the \cppcode{TransformOpInterface} and is easily composable with other transform operations. All facilities for managing the interpreter state and recursively entering the blocks are available in the same way as they are for “regular” transform operations. Match operations are expected to return a silenceable failure to indicate failure to match, and to immediately propagate definite failures. If they have nested operations, they are expected to handle and, in most cases, silence the silenceable failures produced when applying those operations. For our operation, the matching is essentially a loop iterating over all operands of the (single) payload operation and applying nested transform ops until they all succeed for one of the operands.

\begin{cppblock}
// Matcher ops implement `apply` similarly to other transform ops. They are not
// expected to modify payload, but use the tri-state result to signal failure or
// success to match, as well as potential irrecoverable errors.
mlir::DiagnosedSilenceableFailure
mlir::transform::HasOperandSatisfyingOp::apply(
    mlir::transform::TransformRewriter &rewriter,
    mlir::transform::TransformResults &results,
    mlir::transform::TransformState &state) {
  // For simplicity, only handle a single payload op. Actual implementations
  // can use `SingleOpMatcher` trait to simplify implementation and document
  // this expectation.
  auto payloadOps = state.getPayloadOps(getOp());
  if (!llvm::hasSingleElement(payloadOps))
    return emitSilenceableError() << "expected single payload";

  // Iterate over all operands of the payload op to see if they can be matched
  // using the body of this op.
  Operation *payload = *payloadOps.begin();
  for (OpOperand &operand : payload->getOpOperands()) {
    // Create a scope for transform values defined in the body. This corresponds
    // to the syntactic scope of the region attached to this op. Any values
    // associated with payloads from now on will be automatically dissociated
    // when this object is destroyed, i.e. at the end of the iteration.
    // Associate the block argument handle with the operand.
    auto matchScope = state.make_region_scope(getBody());
    if (failed(state.mapBlockArgument(getBody().getArgument(0),
                                      {operand.get()}))) {
      return DiagnosedSilenceableFailure::definiteFailure();
    }

    // Iterate over all nested matchers with the current mapping and see if they
    // succeed.
    bool matchSucceeded = true;
    for (Operation &matcher : getBody().front().without_terminator()) {
      // Matcher ops are applied similarly to any other transform op.
      DiagnosedSilenceableFailure diag =
          state.applyTransform(cast<TransformOpInterface>(matcher));

      // Definite failures are immediately propagated as they are irrecoverable.
      if (diag.isDefiniteFailure())
        return diag;

      // On success, keep checking the remaining conditions.
      if (diag.succeeded())
        continue;

      // Report failure-to-match for debugging purposes and stop matching this
      // operand.
      assert(diag.isSilenceableFailure());
      DEBUG_MATCHER(DBGS_MATCHER()
                    << "failed to match operand #" << operand.getOperandNumber()
                    << ": " << diag.getMessage());
      (void)diag.silence();
      matchSucceeded = false;
      break;
    }
    // If failed to match this operand, try other operands.
    if (!matchSucceeded)
      continue;

    // If we reached this point, the matching succeeded for the current operand.
    // Remap the values associated with terminator operands to be associated
    // with op results, and also map the parameter result to the operand's
    // position. Note that it is safe to do here despite the end of the scope
    // as `results` are integrated into `state` by the interpreter after `apply`
    // returns rather than immediately.
    SmallVector<SmallVector<MappedValue>> yieldedMappings;
    transform::detail::prepareValueMappings(
        yieldedMappings, getBody().front().getTerminator()->getOperands(),
        state);
    results.setParams(getPosition().cast<OpResult>(),
                      {rewriter.getI32IntegerAttr(operand.getOperandNumber())});
    for (auto &&[result, mapping] : llvm::zip(getResults(), yieldedMappings))
      results.setMappedValues(result, mapping);
    return DiagnosedSilenceableFailure::success();
  }

  // If we reached this point, none of the operands succeeded the match.
  return emitSilenceableError()
         << "none of the operands satisfied the conditions";
}
\end{cppblock}

By convention, operations implementing \cppcode{MatchOpInterface} must not modify payload IR and must therefore specify that they only read operand handles and payload as their effects.

\begin{cppblock}
void transform::CollectMatchingOp::getEffects(
    SmallVectorImpl<MemoryEffects::EffectInstance> &effects) {
  onlyReadsHandle(getRoot(), effects);
  producesHandle(getResults(), effects);
  onlyReadsPayload(effects);
}
\end{cppblock}

This operation can now be included in a transform dialect extension, loaded and used in our matcher. Specifically, we will use it to indicate that either of the operands of the “max” elementwise operation in our example can be produced by the previous elementwise operation. The previous operation will still require the matmul to produce the first operand for simplicity. The updated matcher sequence looks as follows.

\begin{mlirblock}
transform.named_sequence @match_matmul_elemwise(
    %last: !transform.any_op {transform.readonly})
    -> (!transform.any_op, !transform.any_op, !transform.any_op,
        !transform.param<i32>) {
  // The last operation must be an elementwise binary.
  transform.match.operation_name %last ["linalg.elemwise_binary"]
    : !transform.any_op

  // One of its operands must be defined by another operation, to which we
  // will get a handle here. This is achieved thanks to a newly defined
  // operation that tries to match operands one by one using the match
  // operations nested in its region.
  %pos, %middle = transform.match.my.has_operand_satisfying %last
      : (!transform.any_op) -> (!transform.param<i32>, !transform.any_op) {
  ^bb0(%operand: !transform.any_value):
    // The operand must be defined by an operation.
    %def = transform.get_defining_op %operand
      : (!transform.any_value) -> !transform.any_op
    // The defining operation must itself be an elementwise binary.
    transform.match.operation_name %def ["linalg.elemwise_binary"]
      : !transform.any_op
    transform.yield %def : !transform.any_op
  }

  // And the first operand of that operation must be defined by yet another
  // operation.
  %matmul = transform.get_producer_of_operand %middle[0]
    : (!transform.any_op) -> !transform.any_op
  // And that operation is a matmul.
  transform.match.operation_name %matmul ["linalg.matmul"] : !transform.any_op
  // We will yield the handles to the matmul and the two elementwise
  // operations separately.
  transform.yield %matmul, %middle, %last, %pos
    : !transform.any_op, !transform.any_op, !transform.any_op,
      !transform.param<i32>
}
\end{mlirblock}

This achieves the desired effect and matches both \mintinline{text}{max(add(matmul(...), bias), 0)} and \mintinline{text}{max(0, add(matmul(...), bias))} in the same values. The \mlircode{

In order to demonstrate that groups of operations are matched independently of each other, let us use the \mlircode{transform.foreach_match} operation that allows one to implement a simple high-level pattern rewriting approach within the transform dialect (for advanced or lower-level pattern rewriting, consider PDL(L) or C++ rewriting APIs). It maps a matcher named sequence to an action named sequence, and the latter gets invoked whenever the former succeeds.

\begin{mlirblock}
// Traverses the payload IR associated with the operand handle, invoking
// @match_matmul_elemwise on each of the operations. If the named sequence
// succeeds, i.e., if none of the nested match (transform) operations
// produced a silenceable failure, invokes @print_matmul_elemwise and
// forwards the values yielded as arguments of the new invocation. If the
// named sequence fails with a silenceable failure, silences it (the message
// is forwarded to the debug stream). Definite failures are propagated
// immediately and unconditionally, as usual.
transform.foreach_match in %root
  @match_matmul_elemwise -> @print_matmul_elemwise
  : (!transform.any_op) -> !transform.any_op

\end{mlirblock}
The \mlircode{@print_matmul_elemwise} named sequence, available in \mintinline{text}{multiple.mlir}, will use the parameter with the position of the operand to differentiate the two groups.

\subsection{Matchers for Inferred Features}
The matcher sequences described above, although useful to drive transformations from within the transform dialect interpreter, are rather basic since they mostly rely on operation names and use-def chains. Alternative implementations using APIs or various declarative rewrite rules are barely less expressive and sometimes more concise. The real power of transform dialect matcher ops lies in the possibility to define matchers of inferred properties of payloads, i.e., properties that are not directly accessible as an attribute of an operation or any straightforward relation between IR components.

The utility of such matchers can be easily demonstrated by slightly modifying our original example. If matrix multiplication is expressed as a special case of tensor contraction using \mlircode{linalg.generic} instead of \mlircode{linalg.matmul}, the operation name-based matcher no longer applies. Yet such a representation is very common and can appear both in the original input and during the course of transformation, e.g., where a higher-dimensional contraction is decomposed into loops around a matrix multiplication.

In order to be a (potentially transposed) matrix multiplication, the \mlircode{linalg.generic} operation must have the following features:

\begin{itemize}
\item Total rank of 3.
\item Two inputs accessed as projected permutation of iteration dimensions.
\item One output accessed as projected permutation of iteration dimensions.
\item Iteration dimensions can be subdivided into LHS parallel, RHS parallel and reduction dimensions.
\item The body block consists of a multiplication and an addition.
\end{itemize}

Most of these features can be derived from the properties of the operation, e.g., the total rank corresponds to the number of entries in the iterators attribute, but almost none of them are immediately accessible in the IR or in any declarative form, which is usually limited to checking the presence or the exact match of an attribute or a type. The transform dialect allows these features to be implemented in the apply method of a matcher op and reused across multiple matching cases. For structured linear algebra payload operations, many such match operations are readily available in the structured extension. They are sufficient to implement a matrix multiplication matcher using the features listed above almost verbatim.

\begin{mlirblock}
transform.named_sequence @match_generic_matmul(
    %candidate: !transform.any_op {transform.readonly}) -> !transform.any_op {
  // Match a structured linear algebra operation.
  transform.match.structured %candidate : !transform.any_op {
  ^bb0(%c: !transform.any_op):
    // With a rank equal to 3.
    %rank = transform.match.structured.rank %c
      : (!transform.any_op) -> !transform.param<i64>
    %c3 = transform.param.constant 3 : i64 -> !transform.param<i64>
    transform.match.param.cmpi eq %rank, %c3 : !transform.param<i64>

    // With 2 inputs.
    %n_ins = transform.match.structured.num_inputs %c
      : (!transform.any_op) -> !transform.param<i64>
    %c2 = transform.param.constant 2 : i64 -> !transform.param<i64>
    transform.match.param.cmpi eq %n_ins, %c2 : !transform.param<i64>

    // With 1 output (note that structured ops in destination passing style
    // has as many inits as outputs).
    %n_inits = transform.match.structured.num_inits %c
      : (!transform.any_op) -> !transform.param<i64>
    %c1 = transform.param.constant 1 : i64 -> !transform.param<i64>
    transform.match.param.cmpi eq %n_inits, %c1 : !transform.param<i64>

    // All inputs and inits are accessed with a projected permutation.
    transform.match.structured.input %c[all] {projected_permutation}
      : !transform.any_op
    transform.match.structured.init %c[0] {projected_permutation}
      : !transform.any_op

    // The body is a mulf/addf contraction with appropriate dimensions.
    transform.match.structured.body %c
      { contraction = ["arith.mulf", "arith.addf"] } : !transform.any_op
    %batch, %lhs, %rhs, %reduction =
    transform.match.structured.classify_contraction_dims %c
      : (!transform.any_op)
      -> (!transform.param<i64>, !transform.param<i64>, !transform.param<i64>,
          !transform.param<i64>)


    // There is one of lhs, rhs and reduction dimensions and zero batch
    // dimensions.
    %n_batch = transform.num_associations %batch
      : (!transform.param<i64>) -> !transform.param<i64>
    %n_lhs = transform.num_associations %lhs
      : (!transform.param<i64>) -> !transform.param<i64>
    %n_rhs = transform.num_associations %rhs
      : (!transform.param<i64>) -> !transform.param<i64>
    %n_reduction = transform.num_associations %reduction
      : (!transform.param<i64>) -> !transform.param<i64>
    %c0 = transform.param.constant 0 : i64 -> !transform.param<i64>
    transform.match.param.cmpi eq %n_batch, %c0 : !transform.param<i64>
    transform.match.param.cmpi eq %n_lhs, %c1 : !transform.param<i64>
    transform.match.param.cmpi eq %n_rhs, %c1 : !transform.param<i64>
    transform.match.param.cmpi eq %n_reduction, %c1 : !transform.param<i64>
  }
  transform.yield %candidate : !transform.any_op
}
\end{mlirblock}

While this example leverages the contraction-specific matchers that have a rather non-trivial C++ implementation, the transform dialect is sufficiently flexible to implement this reasoning directly if desired. One could, for example, obtain the access map of each input as a parameter and extract the accessed dimensions as other parameters that can be compared with each other to ensure the subscripts are m,k for LHS, k,n for RHS and m,n for the init/result given the m,n,k notation for loops.

\appendix

\section{Reproducing Halide Schedule}

This chapter demonstrates how a schedule from the
\href{http://halide-lang.org}{Halide DSL}~\cite{halide} can be implemented
using Transform dialect for structured ops.

Note that the IR below is pseudo-code with types removed for brevity. It may
also get out of sync with the current syntax. Always refer to the source code
in \mintinline{text}{mlir/examples/transform/ChH} as the source of truth.

\subsection{Channeled Convolution}

The Transform dialect provides a substrate for implementing “transformation
directive” domain-specific languages (DSLs) in MLIR. Such a DSL, at least in its
scheduling part, can target the operations in the Transform dialect that are
later applied by the compiler. Sets of transform operations, or even new
dialects leveraging the same interfaces and infrastructure, can be added to
support a specific DSL for a particular scheduling model. In this chapter, we
will revisit the Halide DSL that has (re)popularized separate specification of
schedules originally for image processing programs.

Two approaches Halide to the Transform dialect are possible:

\begin{itemize}
\item  Create a new dialect that corresponds to the computational part of Halide
  DSL, and define a set of transformations wrapped into Transform dialect
  operations, that correspond to the scheduling part of the DSL.
\item Map the Halide abstractions to the existing MLIR abstractions, for both
  parts of the DSL.
\end{itemize}

We will consider the latter approach as the computational part of the DSL easily
maps to the structured ops in the Linalg dialect. This also gives us the
opportunity to discuss how Linalg transformations on the so-called structured
operations are similar to or different from the existing transformations.

We will consider the 2D channeled convolution example extracted from Halide
\href{https://github.com/halide/Halide/tree/294f80c49bf3bb8582446613c25fcce03b82bcd8/apps/conv_layer}{application
examples}.

\begin{cppblock}
// Sizes of the problem.
const int N = 5, CI = 128, CO = 128, W = 100, H = 80;

// Sized inputs. Note that the order of dimensions is
// inverted in Halide with respect to C++, so the last dimension
// in the list (N for input, CI for filter) is the least
// frequently varying. The C++ equivalent is input[N][H+2][W+2][CI].
Buffer<float, 4> input({CI, W+2, H+2, N}, "input");
Buffer<float, 4> filter({CO, 3, 3, CI}, "filter");
Buffer<float, 1> bias(std::vector<int>{CO}, "bias");

// ... data initialization happens here ...

// Declarations of "mathematical functions" for convolution and relu.
Func conv("conv"), relu("relu");

// Iterators/subscripts.
Var x("x"), y("y"), c("c"), n("n");

// 3D reduction domain (channels and 2 window dimensions),
// dimensions are later referred to as r.x, r.y, r.z.
RDom r(0, CI, 0, 3, 0, 3);

// Core convolution with the result initialized to the bias value.
// Note that the order of iterators is inverted in Halide DSL,
// i.e. `n` corresponds to the lest frequently-varying (outermost) dimension
// here and below.
conv(c, x, y, n) = bias(c);
conv(c, x, y, n) += filter(c, r.y, r.z, r.x) * input(r.x, x + r.y, y + r.z, n);

// ReLU rectification, an elementwise operation.
relu(c, x, y, n) = max(0, conv(c, x, y, n));
\end{cppblock}

This can be almost directly converted to Linalg dialect operating on tensors,
which is conceptually closer to the ``mathematical function'' abstraction and
is where the majority of transformations are available.

\begin{mlirblock}
// Bias. Using a named Linalg operation for brevity.
%bias_init = tensor.empty() : !toutput
%biased = linalg.broadcast ins(%bias : !tbias)
                          outs(%bias_init : !toutput) dimensions = [0, 1, 2]

// Convolution proper. While Linalg has named operations for 2D convolutions,
// the one in the Halide example has an uncommon order of filter dimensions
// and is not supported. It also takes the filter as first argument. This
// code recreates it faithfully using the generic form.
%convolved = linalg.generic {
  iterator_types = ["parallel", "parallel", "parallel", "parallel",
                    "reduction", "reduction", "reduction"],
  indexing_maps = [
    affine_map<(n, y, x, c, rz, ry, rx) -> (rx, rz, ry, c)>,
    affine_map<(n, y, x, c, rz, ry, rx) -> (n, y+rz, x+ry, rx)>,
    affine_map<(n, y, x, c, rz, ry, rx) -> (n, y, x, c)>
  ]
} ins(%filter, %input: !tfilter, !tinput)
  outs(%biased : !toutput) {
^bb0(%in: f32, %f: f32, %b: f32):
  // Note the fastmath attributes that allow operations to be recombined into
  //   %0 = math.fma %in, %f, %b : f32
  // later on and to reorder reductions.
  %m1 = arith.mulf %in, %f  {fastmath = #arith.fastmath<fast>} : f32
  %0 = arith.addf %b, %m1  {fastmath = #arith.fastmath<fast>} : f32
  linalg.yield %0 : f32
} -> !toutput

// ReLU is just a max(0, x).
%c0 = arith.constant 0.0 : f32
%relued = linalg.generic {
  iterator_types = ["parallel", "parallel", "parallel", "parallel"],
  indexing_maps = [
    affine_map<(d0, d1, d2, d3) -> ()>,
    affine_map<(d0, d1, d2, d3) -> (d0, d1, d2, d3)>,
    affine_map<(d0, d1, d2, d3) -> (d0, d1, d2, d3)>
  ]
} ins(%c0, %convolved : f32, !toutput)
  outs(%output : !toutput) {
^bb0(%cst: f32, %in: f32, %out: f32):
  %0 = llvm.intr.maxnum(%cst, %in) : (f32, f32) -> f32
  linalg.yield %0 : f32
} -> !toutput
\end{mlirblock}

In Halide, a function such as \mlircode{conv} may consist of two parts: a ``functional''
initialization computation and an in-place update for reductions. This is
expressed as two C++ statements in the embedded DSL, but internally is
represented in a single object. Linalg doesn’t have such a capability to the
initialization and the update are represented as two distinct Linalg operations
that are not connected to each other. Furthermore, the \mintinline{text}{x, y, c, n}
variables in Halide DSL correspond to implicit loops iterating over the
corresponding objects, which implies that functions sharing these variables in
their definitions also share the corresponding loops. In other words, the loop
equivalent of the Halide definition starts in a fully-fused form. The Linalg
model is the opposite with each structured operation corresponding to its own
loop nest, resulting in a fully-distributed form. This will affect how the
schedule is constructed later on.

The loop structure for Halide computation resembles the following (adapted from
debug dump with \mintinline{shell}{HL_DEBUG_CODEGEN=1})

\begin{pyblock}
for n
  for y
    for x
      for c
        conv[n, y, x, c] = bias[c]
        for rz
          for ry
            for rx
              conv[n, y, x, c] += filter[rx, rz, ry, c] * input[n, y+rz, x+ry, rx]
        relu[n, y, x, c] = max(0, conv[n, y, x, c])
\end{pyblock}

The loop structure for the Linalg computation is as follows (obtained by
\mintinline{text}{mlir-opt}\\ \mintinline{text}{--linalg-generalize-named-ops --empty-tensor-to-alloc-tensor --one-shot-bufferize}\\ \mintinline{text}{--convert-linalg-to-loops})

\begin{pyblock}
for n
  for y
    for x
      for c
        init[n, y, x, c] = bias[c]
for n
  for y
    for x
      for c
        for rz
          for ry
            for rx
              conv[n, y, x, c] += filter[rx, rz, ry, c] * input[n, y+rz, x+ry, rx]
for n
  for y
    for x
      for c
        relu[n, y, x, c] = max(0, conv[n, y, x, c])

\end{pyblock}

\subsection{Mapping Halide Scheduling Primitives to Linalg Structured Transforms}

The complete Halide schedule listed in the example is as follows

\begin{cppblock}
Var co, ci, xo, xi;
relu.split(c, co, ci, vec * tile_w)
  .split(x, xo, xi, tile_h)
  .reorder(ci, xi, xo, y, n, co)
  .vectorize(ci, vec)
  .unroll(ci)
  .unroll(xi)
  .parallel(y)
  .parallel(n)
  .parallel(co);

conv.compute_at(relu, xo)
  .vectorize(c, vec)
  .unroll(c)
  .unroll(x)
  .unroll(y)
  .update()
  .reorder(c, x, y, r.x, r.y, r.z, n)
  .vectorize(c, vec)
  .unroll(c)
  .unroll(x)
  .unroll(y)
  .unroll(r.x, 2);
\end{cppblock}

We will consider only the case without parallelization to avoid the difference
in parallel runtimes generated by Halide and used by MLIR. This schedule
corresponds to a sequence of loop manipulations, unrolling and vectorization.
The following directives are present and can be mapped to transformations on
Linalg as described below.

\begin{itemize}
\item \mlircode{split} decomposes a loop dimension into two immediately nested loops with
  the inner loop having at most the given number of iterations. This can be
  understood as loop \emph{strip-mining} or a degenerate case of tiling a single
  dimension using any of \mlircode{linalg.tile_} transform ops. We will be using
  \mlircode{transform.structured.tile_using_forall} as this kind of loop is best
  supported by bufferization and can also be turned into a parallel loop later
  on. Unlike Halide, this doesn’t add new dimensions to the original
  operation, but rather creates a loop around it and rewrites the operation
  itself to operate on a subset of the original data.
\item \mlircode{reorder} rearranges the loops arbitrarily. In Linalg representation, loops
  are implicit and are intended to remain so as long as possible to target
  microkernels. The order of implicit loops in a \mlircode{linalg.generic} operation
  can be changed by using \mlircode{transform.structured.interchange}, but this does
  not apply to named operations that need to be “generalized” first by calling
  \mlircode{transform.structured.generalize}. However, this can only reorder implicit
  dimensions and not the explicit loops materialized by tiling operations that
  can no longer be “folded” into the original operation. Instead, we can
  leverage this behavior by materializing loops directly in the desired order
  by “tiling” to size 1.
\item \mlircode{vectorize} indicates that the given dimension should be vectorized with the
  given factor; if the loop extent is larger than the factor, the loop is
  effectively split into two parts and the inner one is vectorized. On the
  contrary, structured Linalg op vectorization applies as a global
  transformation to all suitable operations at, e.g., a function scope via
  \mlircode{transform.structured.vectorize_children_and_apply_patterns}. It relies on
  MLIR’s support for multidimensional vectors to directly map multidimensional
  tensors, which are later decomposed into operations on smaller
  hardware-compatible vectors during lowering.
\item \mlircode{unroll} unrolls the loop fully or up to the given factor. It is
  equivalent to \mlircode{transform.loop.unroll}.
\item \mlircode{compute_at} indicates that the value of the function must be computed
  within the given loop that will be produced for another function; depending
  on the relation between loops surrounding functions, this corresponds to
  either a loop distribution or a producer/consumer fusion. Given that the
  Linalg representation starts in the fully distributed form, it can be
  represented as a sequence of \mlircode{transform.structured.fuse_into_containing_op}
  that operates on \mlircode{forall} loops materialized by tiling beforehand.
\end{itemize}

\subsection{Recreating the Loop Structure}

The three first transformation directives for \mlircode{relu} in the Halide
schedule aim at producing the following loop structure.

\begin{pyblock}
for co
  for n
    for y
      for xo
        for xi
          for ci
            relu[n, y, xo*tile_h + xi, co*tile_w*vec + ci] = ...
\end{pyblock}

Note that the outer part of the \mintinline{text}{c} gets hoisted from all of the surrounding
loops. The implicit loop order for the operation is \mintinline{text}{n, y, x, c}, so the \mintinline{text}{co}
loop needs to be materialized first in order to achieve the desired reordering.
The remaining dimensions can be materialized as loops in one transformation.

\begin{mlirblock}
    //                                                             [n  y  x  c]
    %co, %relu2 = transform.structured.tile_using_forall %relu
                                                        tile_sizes [0, 0, 0, 64]
    %n_y_xo, %relu3 = transform.structured.tile_using_forall %relu2
                                                        tile_sizes [1, 1, 5, 0]
\end{mlirblock}

This will result in the following loops being created in the IR with the nested
elementwise operation operating on a smaller subset of original data via
implicit loops.

\begin{mlirblock}
scf.forall (%co) in (2) {
  scf.forall (%n, %y, %xo) in (5, 80, 20) {
    tensor.extract_slice
    // Implicit dimensions [ni=0:1, y=0:1, xi=0:5, ci=0:64]
    %relued = linalg.elemwise_binary { fun = #linalg.binary_fn<max_signed> } // ...
    scf.forall.in_parallel {
      tensor.parallel_insert_slice // ...
    }
  }
}
\end{mlirblock}

The following loop restructuring transformations are \mintinline{text}{compute_at} and \mintinline{text}{reorder}
on the \mintinline{text}{conv} function that need to happen before loops are destroyed by
unrolling and vectorization. They intend to produce the final desired loop
structure.

\begin{mlirblock}
for co
  for n
    for y
      for xo
        for xi
          for ci
            conv[n, y, x*tile_h + xi, co*tile_w*vec + ci] = ...
        for rz
          for ry
            for rx
              for xi
                for ci
                  conv[n, y, x*tile_h + xi, co*tile_w*vec + ci] += ...
        for xi
          for ci
            relu[n, y, xo*tile_h + xi, co*tile_w*vec + ci] = ...
\end{mlirblock}

Practically, this corresponds to fusing the convolution initialization and
update into the \mintinline{text}{co, n, y, xo} loops materialized by tiling earlier. Structured
op transformation set supports fusing the producer of a value into its consumer,
so fusion happens in two stages:

\begin{itemize}
\item first the main convolution update is fused into ReLU that uses it and has
    loops materialized;
\item then the bias initialization is fused into the convolution+relu loop nest.
\end{itemize}

Each stage consists of two transformations fusing the computational operation
into the outer loop, then the inner loop.

\begin{mlirblock}
%conv2, %co2 = transform.structured.fuse_into_containing_op %conv into %co
%conv3, %n_y_xo2 = transform.structured.fuse_into_containing_op %conv2
  into %n_y_xo

%bias2, %co3 = transform.structured.fuse_into_containing_op %bias into %co2
%bias3, %n_y_xo3 = transform.structured.fuse_into_containing_op %bias2
  into %n_y_xo2
\end{mlirblock}

To complete the structure, we need to put the \mintinline{text}{rz, ry, rx} loops outside the
``tile'' loops \mintinline{text}{xi, ci}. This can be achieved materializing the corresponding
loops from the convolution operation. However, these are reduction loops and it
wouldn’t be valid to materialize them as intrinsically parallel ``forall'' loops.
Instead, we use the dedicated “reduction tiling” transformation and produce
sequential \mintinline{text}{scf.for} loops. (\mlircode{scf.forall} loops can also express parallel
reductions, but the corresponding transformation doesn’t handle reductions along
more than one dimension at the moment of writing.)

\begin{mlirblock}
%rz_ry_rx, %red_fill, %conv4, %comb
  = transform.structured.tile_reduction_using_for %conv3
//               n  y  x  c  rz ry rx
  by tile_sizes=[0, 0, 0, 0, 1, 1, 1]
\end{mlirblock}

This transformation materializes the desired loops around the convolution
operation. It is also more capable than merely producing (reduction) loops: the
transformed code performs \mintinline{text}{tile_size} partial reductions of \mintinline{text}{N / tile_size}
elements, potentially in parallel by changing the dimension kind of the
structured operation inside the loop, and then performs a final reduction of
these partial results by producing a new “combiner” structured operation after
the loops. In our case, \mintinline{text}{tile_size = 1} along all dimensions, so the reduction
is entirely performed by the generated loops. The combiner structured operation
is still produced and adds up the reduction result with the initial value. This
changes the order of floating point operations (so would reduction tiling with
non-unit size) and may affect the final result due to non-commutativity of these
operations, but is explicitly allowed by \mintinline{text}{fastmath} flags. Halide also emits
LLVM IR with full \mintinline{text}{fastmath} flags.

Finally, we need to produce innermost loops \mintinline{text}{xi} and \mintinline{text}{ci} that are still not
explicit. As our next step is going to be vectorization along \mintinline{text}{ci}, we need to
take into account the way it operates on MLIR structured operations: rather than
selecting a specific vector size and loop/dimension to vectorize, it directly
substitutes multidimensional vector types for tensor types and updates the
operations accordingly. Therefore, our tensor type should not become trivial,
i.e. size-1, and retain a \mintinline{text}{vector_size} sized dimension along the desired axis,
\mintinline{text}{ci}. This can be achieved by tiling with \mintinline{text}{vector_size} as tile size in that
dimension:

\begin{mlirblock}
//                                                                  n  y  xi ci
%1, %c5 = transform.structured.tile_using_forall %conv4 tile_sizes [0, 0, 1, 16]
%2, %b4 = transform.structured.tile_using_forall %bias3 tile_sizes [0, 0, 1, 16]
%3, %r4 = transform.structured.tile_using_forall %relu3 tile_sizes [0, 0, 1, 16]
%4, %c2 = transform.structured.tile_using_forall %comb  tile_sizes [0, 0, 1, 16]
\end{mlirblock}

Note that the combiner operation produced by reduction tiling is also tiled here.

\subsection{Explicit Loop Unrolling}

The remaining unhandled loop transformation is unrolling. Specifically,
unrolling is requested for the innermost loops that form the 4x5 tile of
16-element vector operations to ensure a contiguous sequence of \mintinline{text}{vfma}
instructions using 20 512-bit vector registers as accumulators. Unrolling
additional loops,, \mintinline{text}{unroll(y)} and \mintinline{text}{unroll(r.x, 2)}, is requested in the
schedule but \emph{has no practical effect}. That is, the code, and all intermediate
representations, produced by Halide with these directives removed is \emph{strictly
identical} to the code with the full schedule. Therefore, we will only unroll
the corresponding loops corresponding to \mintinline{text}{xi} and \mintinline{text}{ci} dimensions that actually
get unrolled by Halide.

As tiling in the Transform dialect produces handles to the loops materialized by
tiling, unrolling those loops is just a matter of chaining the corresponding
transformation. Note that the inner loop must be unrolled first as unrolling the
outer loop will invalidate the handles to the inner loop.

\begin{mlirblock}
transform.loop.unroll %bias_ci {factor = 4}
transform.loop.unroll %bias_xi {factor = 5}
transform.loop.unroll %conv_ci {factor = 4}
transform.loop.unroll %conv_xi {factor = 5}
transform.loop.unroll %relu_ci {factor = 4}
transform.loop.unroll %relu_xi {factor = 5}
transform.loop.unroll %comb_ci {factor = 4}
transform.loop.unroll %comb_xi {factor = 5}
\end{mlirblock}

\subsection{Vectorization}

These transformations produced the desired loop structure and we are now ready
to vectorize. Before proceeding it is desirable to simplify the code as tiling
and fusion may have produced a lot of operations computing tensor subsets and
loop ranges, some of which may be duplicated or excessively complex.
Simplification involving canonicalization, common subexpression elimination,
loop invariant code motion and various rewrite patterns can be applied directly
from the transform dialect. Furthermore, an arbitrary combination of rewrite
patterns can be applied \emph{in one sweep} to a given scope, a functionality
that \emph{cannot be achieved with conventional compiler passes} that apply
each group of patterns separately (at least without creating a new pass for
each combination of pattern groups).

\begin{mlirblock}
%f00 = transform.structured.match ops{["func.func"]} in %arg0
transform.apply_patterns to %f00 {
  transform.apply_patterns.canonicalization
  transform.apply_patterns.linalg.tiling_canonicalization
}
transform.apply_cse to %f00

%all_loops = transform.structured.match interface{LoopLikeInterface} in %arg0
transform.apply_licm to %all_loops
\end{mlirblock}

One final simplification is necessary to produce good vectorized code.
Tiling-by-one as a way of materializing loops produced structured
(\mlircode{linalg}) operations processing 4D types where only one dimension
isn’t unit-sized, e.g., \mlircode{tensor<1x1x1x16xf32>} where 16 is the vector
size corresponding to AVX512, as structured tiling doesn’t modify the rank of
the operation in order to preserve the original structure. Even though the core
computation is the same, the produced code may end up more complicated than
necessary, in particular when decomposing multidimensional vectors into
single-dimensional vectors supported by hardware. Such unit dimensions can be
explicitly folded away using the corresponding pattern set before
vectorization.

\begin{mlirblock}
transform.apply_patterns to %f00 {
  transform.apply_patterns.linalg.fold_unit_extent_dims_via_reshapes
}

%fv = transform.structured.vectorize_children_and_apply_patterns %f00
\end{mlirblock}

This produces the desired code performing arithmetic operations on
\mlircode{vector<16xf32>} types that can be easily lowered to AVX512 instructions by the
downstream compiler. Vectorization may have created new opportunities for code
simplification, in particular combining tensor subsetting and vector slicing
operations. Another round of simplification can be applied post vectorization.

\begin{mlirblock}
transform.apply_patterns to %fv {
  transform.apply_patterns.canonicalization
  transform.apply_patterns.tensor.fold_tensor_subset_ops_into_vector_transfers
}
transform.apply_cse to %fv
transform.structured.hoist_redundant_vector_transfers %fv
\end{mlirblock}

\subsection{Lowering to LLVM and The Bufferization Hurdle}

With the loop restructuring done, the program now needs to be converted to the
executable form. The first step in doing so is \emph{bufferization}, the
process that associates a memory buffer with every tensor in the payload IR.
MLIR’s one-shot bufferization is directly available as a transform operation.

\begin{mlirblock}
%arg1 = transform.bufferization.one_shot_bufferize %arg0 {
  bufferize_function_boundaries = true,
  function_boundary_type_conversion = 1 : i32 }
\end{mlirblock}

One-shot bufferization itself does not produce buffer deallocations, which may
lead to leaks. So we have to run the buffer deallocation pass pipeline to avoid
them. Note that the Transform dialect seamlessly runs named passes and pass
pipelines: if desired, one could replace complex
\mintinline{text}{--pass-pipeline expressions} with operations. Note that we
apply the pipeline to functions rather than entire module to avoid running it
on the transform IR that is contained in the module.

\begin{mlirblock}
%f = transform.structured.match ops{["func.func"]} in %arg1
  : (!transform.any_op) -> !transform.any_op
transform.apply_registered_pass "buffer-deallocation-pipeline" to %f
  : (!transform.any_op) -> !transform.any_op
\end{mlirblock}

In this particular case, the transformed IR could be directly bufferized. This
is not always the case in general as some operations, in particular
\mlircode{tensor.empty} may not be bufferizable. Such operations need to be
removed before running the bufferization, which can often be achieved by
sufficient fusion (as in our case), or by running
\mlircode{transform.bufferization.eliminate_empty_tensors} that removes the
\mlircode{tensor.empty} operations only serving for defining the size of a
computation or \mlircode{transform.bufferization.empty_tensor_to_alloc_tensor}
that materializes a new temporary buffer for empty tensors to be used as local
caches.

\begin{mlirblock}
// Apply general canonicalization and CSE to each function after
// bufferization as new simplification opportunities may have appeared.
%fb = transform.structured.match ops{["func.func"]} in %arg1
transform.apply_patterns to %fb {
  transform.apply_patterns.canonicalization
}
transform.apply_cse to %fb

// Lower complex, multidimensional vector operations into simpler
// primitives. This particular selection of the pattern groups corresponds
// to vector dialect operations present in the payload IR at this stage.
// Many of these groups can be parameterized to use different strategies or
// lower-level primitives offering performance trade-offs. In this case, we
// are selecting the simplest strategies.
transform.apply_patterns to %fb {
  transform.apply_patterns.vector.lower_contraction
    lowering_strategy = parallelarith
  transform.apply_patterns.vector.lower_transfer
    max_transfer_rank = 1
  transform.apply_patterns.vector.lower_transpose
    lowering_strategy = eltwise
  transform.apply_patterns.vector.lower_shape_cast
}

// These patterns apply in a separate sweep to avoid transfer-to-scf
// patterns overlap with lower-transfer patterns as they apply to the same
// kind of operations. These patterns may produce local allocations to act
// as temporary caches deep inside loops, which could lead to catastrophic
// performance. Such allocations are moved onto the stack and hoisted from
// all the surrounding loops.
transform.apply_patterns to %fb {
  transform.apply_patterns.vector.transfer_to_scf
  transform.apply_patterns.memref.alloc_to_alloca
  }
transform.bufferization.buffer_loop_hoisting %fb

// A final round of cleanups additionally includes patterns to simplify
// buffer aliasing operations that may have been introduced during
// bufferization and could result in excessively complex address
// computation.
transform.apply_patterns to %fb {
  transform.apply_patterns.memref.fold_memref_alias_ops
  transform.apply_patterns.canonicalization
}
transform.apply_cse to %fb
\end{mlirblock}

Due to its inter-procedural nature, one-bufferization processes the entire
payload module and thus invalidates all previously created handles. Therefore,
it is typically a late step in the transformation sequence where precise
targeting of transformation is no longer required. The following transformations
are typically module- or function-wide rewrites that are often pattern-based
lowerings. This part of the sequence can be seen as a pass pipeline specified
directly in the transform dialect, with pattern-based lowering passes
constructed \emph{on-the-fly} from named groups of patterns.

The resulting IR can be further completely lowered to the LLVM dialect, then to
LLVM IR and processed by the LLVM compiler to produce an executable or JITted.

The generated code runs in $\approx 420$ ms on an Intel processor with Skylake
microarchitecture clocked at 2.0GHz. Given that the computation performs
$5 \cdot 80 \cdot 100 \cdot 128 \cdot (2 \cdot 3 \cdot 3 \cdot 128 + 2) \approx 5.9 * 10^9$
floating point operations, it reaches $\approx 14$ GFlops. With 1 FMA unit available,
the single-core performance of the test processor is 64 GFlops
($16 \cdot 2 \cdot 2 \cdot 10^9$, where $16$ is the vector width), so only
$22\%$ of the theoretical peak is achieved.

The code produced by Halide runs in $\approx 120$ ms on the same processor, a $3.5 \times$
improvement and $77\%$ of peak. Let us analyze the generated assembly to understand
the source of the difference. The main computational effort is expected to
happen around floating point multiplications and additions in the convolution.
In both cases, the assembly features AVX512 \mintinline{text}{vfma231ps} instructions operating
on \mintinline{text}{
interspersed with memory accesses loading \emph{two} of the \mintinline{text}{fma} operands before
each operation and leading to increased latency.

\begin{asmblock}
vmovups       -192(%r10), %zmm0
vbroadcastss  -1536(%rdi,%r9), %zmm1
vmovups       112(%rsp), %zmm2
vfmadd231ps   %zmm1, %zmm0, %zmm2     # zmm2 = (zmm0 * zmm1) + zmm2
vmovups       %ymm2, 112(%rsp)
vextractf64x4 $1, %zmm2, 144(%rsp)
// 19 more blocks of either
//  (a) vmovups,vbroadcast,vfma(z,z),vextract,
//  (b) vbroadcast,vfma(z,mem),vextract
\end{asmblock}

The Halide-generated code however features compact blocks of \mintinline{text}{vfma231ps} and
\mintinline{text}{vbroadcastss} loading one of the operands while the other two are resident in
registers and loaded before \mintinline{text}{fma}.

\begin{asmblock}
vbroadcastss    -1536(%rsi,%rbx), %zmm25
vmovups         -192(%rdi), %zmm26
vmovups         -128(%rdi), %zmm27
vmovups         -64(%rdi), %zmm28
vmovups         (%rdi), %zmm29
vfmadd231ps     %zmm25, %zmm26, %zmm24  # zmm24 = (zmm26 * zmm25) + zmm24
vfmadd231ps     %zmm25, %zmm27, %zmm23  # zmm23 = (zmm27 * zmm25) + zmm23
vfmadd231ps     %zmm25, %zmm28, %zmm22  # zmm22 = (zmm28 * zmm25) + zmm22
vfmadd231ps     %zmm25, %zmm29, %zmm21  # zmm21 = (zmm29 * zmm25) + zmm21
vbroadcastss    -1024(%rsi,%rbx), %zmm25
vfmadd231ps     %zmm25, %zmm26, %zmm20  # zmm20 = (zmm26 * zmm25) + zmm20
vfmadd231ps     %zmm25, %zmm27, %zmm19  # zmm19 = (zmm27 * zmm25) + zmm19
vfmadd231ps     %zmm25, %zmm28, %zmm18  # zmm18 = (zmm28 * zmm25) + zmm18
vfmadd231ps     %zmm25, %zmm29, %zmm17  # zmm17 = (zmm29 * zmm25) + zmm17
vbroadcastss    -512(%rsi,%rbx), %zmm25

// 3 more blocks of 4 vfmadd231 followed by a vbroadcast
\end{asmblock}

Inspecting the progressive intermediate representations produced by MLIR, one
can observe the \mintinline{text}{load(transfer)}/\mintinline{text}{fma}
interspersing at all levels starting after schedule application. The repeated
tensor subsetting operations, that are later transformed into vector transfer
operations, and vector memory loads, are produced by loop unrolling that was
explicitly requested in the schedule! The issue is the single-assignment model
of tensors (and vectors) that results in long and complex chains of access and
update operations that become so long that the lower-level transformations and
the downstream compiler can no longer simplify them. In fact, unrolling loops
early in the transformation sequence can lead to all sorts of
compiler-performance related problems (including the compiler failing to
perform some optimizations due to excessive code length) in the process.

It is therefore desirable to perform loop unrolling at a later stage,
specifically after bufferization and relevant simplification. However,
bufferization invalidates all loop handles including to loops that we are
willing to unroll. This hurdle can be overcome by matching the payload IR
operations after bufferization to produce new handles. We will first change the
kind of loops produced in the schedule from \mlircode{scf.for} to \mlircode{scf.forall} to have
less operations to match by using \mlircode{transform.structured.tile_using_forall}
instead of \mlircode{transform.structured.tile} when tiling with sizes \mlircode{[0, 0, 1, 16]}.
Then we can match all \mlircode{scf.forall} operations in the payload IR and transform
them into single-iterator \mlircode{scf.for} loops \emph{after bufferization}.

\begin{mlirblock}
%foralls = transform.structured.match ops{["scf.forall"]} in %arg1
%xi_bias, %ci_bias = transform.loop.forall_to_for %xi_ci_bias
%xi_conv, %ci_conv = transform.loop.forall_to_for %xi_ci_conv
%xi_relu, %ci_relu = transform.loop.forall_to_for %xi_ci_relu
%xi_comb, %ci_comb = transform.loop.forall_to_for %xi_ci_comb
\end{mlirblock}

We can then move our loop unrolling transformations later in the transformation
sequence as desired. Compiling this new version to assembly produces exactly the
same core computation around \mintinline{text}{vfmadd231ps} as Halide’s version, which only
differs slightly in allocated registers. Unsurprisingly, this version runs
roughly in 120ms on the same machine.

\subsection{Multi-Dimensional Vectors to the Rescue}

While we managed to produce similar code to Halide in the previous section, we
did so by rematching generated loops after bufferization, which partially defies
the purpose of using handles to chain transformations in the Transform dialect.
Luckily, this step is not really necessary. It only served as an exercise in
producing the desired loop structure.

Multidimensional structured operations on vectors are lowered to target-specific
vectors by unrolling and splitting. For example, an elementwise arithmetic
operation on \mintinline{text}{vector<5x64xf32>} is replaced with 5 operations on
\mintinline{text}{vector<64xf32>} and additional vector value manipulations to recreate the
required type at the MLIR level. Each of these operations is then split into 4
operations on \mintinline{text}{vector<16xf32>} at the LLVM level where the information about
the target vector width becomes available. Collectively, this has exactly the
same effect as first materializing the 5x4 loop nest, and then fully unrolling
these loops. Therefore, the last stage of tiling, re-matching and unrolling can
be removed from the schedule.

The resulting assembly has all \mlircode{vbroadcast} grouped together before \mlircode{vfmadd231}
but otherwise has a similar structure. This grouping is due to each
multi-dimensional vector operation being “unrolled” separately. When executed,
it runs in $\approx 110$ms, a slight improvement of $8\%$ over both the previous version and
Halide, and reaches $\approx 53.7$~GFlop/s or $84\%$ of peak single-core performance. The
improvement is largely due to the intermediate representation being shorter and
simpler in presence of large-vector operations, which allowed for more
aggressive address computation and load placement optimization.

\bibliographystyle{abbrv}
\bibliography{biblio}

\end{document}